\documentclass[fleqn,usenatbib]{mnras}
\usepackage[T1]{fontenc}
\usepackage{ae,aecompl}
\usepackage{graphicx}
\usepackage{amsmath}
\usepackage{amssymb}
\usepackage{natbib}
\usepackage{epstopdf}
\usepackage{lscape}
\usepackage{rotating}
\usepackage{booktabs}

\usepackage{float}

\usepackage{times}
\usepackage{amsmath}
\usepackage{amssymb}
\usepackage{amsfonts}
\usepackage{mathrsfs}
\usepackage{hyperref}
\usepackage{booktabs} 
\usepackage{siunitx}  
\DeclareRobustCommand{\VAN}[3]{#2}
\let\VANthebibliography\thebibliography
\def\thebibliography{\DeclareRobustCommand{\VAN}[3]{##3}\VANthebibliography}

\title [Li-rich M dwarfs]{Lithium-rich M-dwarfs at the ZAMS: Evidence for planetary engulfment?}
\author[R. D. Jeffries et al.]
  {R. D.~Jeffries$^1$, R.~J. Jackson$^1$ and I. Baraffe$^{2,3}$ \\
  $^1$Astrophysics Group, Keele University, Keele, 
      Staffordshire ST5 5BG\\
$^2$ Physics \& Astronomy Department, University of Exeter, Exeter, EX4 4QL, UK\\
$^3$  Ecole Normale Supérieure de Lyon, CRAL, CNRS UMR 5574, 69364, Lyon Cedex 07, France
}

\date{Accepted 26th April 2026}

\pagerange{\pageref{firstpage}--\pageref{lastpage}} \pubyear{2026}

\def\LaTeX{L\kern-.36em\raise.3ex\hbox{a}\kern-.15em
    T\kern-.1667em\lower.7ex\hbox{E}\kern-.125emX}

\begin{document}
\label{firstpage}
\maketitle

\begin{abstract}
We identify 6 early M-dwarfs, in 3 open clusters (NGC~2451a, Blanco~1 and NGC~2516) at ages of 50-200\,Myr, that are anomalously enriched in lithium compared with Li-depleted siblings of similar spectral type. The Li-rich outliers represent 2-3 per cent of stars with $3560 < T_{\rm eff}/{\rm K} < 4045$ in clusters at those ages but are otherwise indistinguishable in their positions, parallaxes and kinematics from other cluster members; their placement in absolute colour-magnitude diagrams is incompatible with being much younger Li-rich interlopers, only one shows evidence of binarity and they are all slow rotators.
The enhanced Li abundances are consistent with the engulfment of 3-10\,$M_\oplus$ of volatile-depleted planetary material after the formation of a radiative core has ended rapid pre main sequence Li depletion. Published planetary formation simulations featuring engulfment via dynamical interactions, and the preponderance of Earth-like exoplanets in close orbits around M-dwarfs, offer some support to this scenario. The observed occurrence rate would be a lower limit to the frequency with which such engulfment events occur between ages of $\sim (30-200)$\,Myr, that depends in the timescale for ongoing Li depletion at the ZAMS.

\end{abstract}


\section{Introduction}

Engulfment of planets or the ingestion of planetary material by a host star is a possible outcome of planetary formation and the evolution of planetary systems \citep[e.g.,][]{Metzger2012a, JacksonPlanet2018}. 
Engulfment might happen as a result of migration processes in the protoplanetary disc \citep{Hasegawa2013a};  dynamical interactions and scattering, either with other planets \citep{Chatterjee2008a} or initiated by close encounters with other stars in clustered environments \citep{Davies2014a}; combined with the dissipative effects of dynamical and equilibrium tides \citep{Gallet2017a, Rao2018a} or magnetic torques \citep{Ahuir2021a}. 

Observational evidence for the frequency and timing of planetary engulfment is difficult to obtain. The accretion of volatile-depleted planetary material is expected to increase the relative abundances of refractory elements. The effects are maximised if the accreted material is mixed into host stars with thin convection zones (CZs). This could lead to small but measurable observational signatures if the photospheric abundances of those stars can be carefully compared with siblings born as part of a multiple system or a cluster, assuming they were born with similar initial abundances. There are numerous studies of individual binary systems with reported small differences in the abundances of their components, suggestive of differential enrichment of refractory elements after formation \citep[e.g.,][]{Saffe2017a, Oh2018a, Jofre2021a, YanaGalarza2024a}. Systematic surveys of wide binary samples claim occurrence rates for these differential refractory enhancements of anywhere between 3 and 35 per cent in solar-type stars and these might be consistent with the ingestion of 2-20\,$M_{\oplus}$ of rocky material \citep{Spina2021a, Behmard2023b, Liu2024a}.

There are both observational challenges and physical uncertainties in the interpretation of these results. The abundance differences are so small, usually $<0.1$ dex, that their secure measurement requires careful differential analyses of high signal-to-noise spectra and preferably, component stars that are twins in terms of their ages, surface gravities and temperatures. The abundances may also be modified by other processes  - atomic diffusion and thermohaline mixing - that can serve to dilute any accretion-related differences on short timescales or even produce differences in binary components with different masses, especially in those with thin CZs where the accretion signatures are hoped to be strongest \citep*{Turcotte1998a, Vauclair2004a, Behmard2023a}. There is also the possibility that inhomogeneity in the ISM and the dust/gas ratio from which these wide binaries formed may contribute some of the differences \citep[e.g.,][]{Hawkins2020a, Soliman2025a}. Finally, the ages of these main sequence binary systems are quite poorly constrained and it is difficult to identify, or test consistency with, specific planetary accretion scenarios.

\subsection{Lithium in M dwarfs as a planetary ingestion signature}

Lithium abundances have also been proposed as a potential planetary ingestion signature \citep{Sandquist2002a, Soares-Furtado2021a}. There have been some observations of Li-enriched stars with relatively thin CZs in clusters \citep{Laws2003a, Ashwell2005a}, in binary systems \citep{Melendez2017a} and in an F-type subgiant hosting a hot Jupiter \citep{Schulte2025}. A complication here is that the dominant $^7$Li isotope is destroyed in stars by $(p,\alpha)$ reactions at relatively low interior temperatures \citep[$\sim 3\times 10^6$\,K,][]{Bildsten1997a} and $^6$Li burns at even lower temperatures. Both convective pre main-sequence (PMS) mixing and subsequent mass- and probably rotation-dependent, non-convective mixing on the main sequence subsequently deplete photospheric $^7$Li (and destroy $^6$Li completely). It is observed that coeval stars in binaries, with and without other planetary ingestion signatures, or in clusters at a range of ages, have a wide dispersion in $^7$Li abundance \citep{Jeffries2023a, Sun2025a}. This makes it difficult to interpret the Li abundance of any individual star with a poorly constrained age, or any Li differences between coeval stars, even of similar mass, as conclusive evidence for accreted material. 

Nevertheless, there is a little-explored combination of stellar mass and ingestion epoch that should lead to very large and unambiguous Li signatures of planetary engulfment. The rapid PMS depletion of Li, facilitated by deep convection in low-mass stars, is in principle halted at a very low Li abundance by the initiation and growth of a substantial radiative core in stars with mass  $\geq 0.5 M_\odot$ \citep[e.g.,][]{Tognelli2011a, Baraffe2015a}. Any subsequent accretion of Li-rich material would be mixed into the still-deep CZ but, if sufficient is accreted, even after dilution it should cause these stars to stand out from coeval stars of similar mass which should be almost entirely Li-depleted. In contrast to other refractory elements, both accretion of volatile-rich (e.g. a close-in giant planet) or volatile-depleted material could produce such a signature, since either would be extremely enriched with Li compared with the existing CZ gas.

We expect a window of opportunity both in mass and time. For stars of low enough mass, the normal expectation would be PMS Li depletion to an undetectably low level. This provides the contrast necessary to detect anomalous enrichment. But the star needs to be of high enough mass that a significant radiative core develops; then, any newly accreted Li can enrich the outer CZ without being immediately burned at the core. The star must also be old enough that rapid PMS Li depletion has ended and the radiative core established, but must be observed on a timescale after any accretion event that is shorter than the timescale for any further Li destruction that would erase the enrichment signature. This scenario has been quantitatively modelled by \cite*{Sevilla2022a}, who predict that a ZAMS engulfment event of $10\,M_\oplus$ of volatile-depleted material could lead to an observable enhancement in photospheric Li abundance lasting a further 10-200\,Myr in 0.5-0.8\,$M_\odot$ stars.

This paper's objective is to search for anomalously Li-rich examples of late-K and early M-dwarfs in open clusters at ages $>30$\,Myr, which should have (i) depleted almost all of their Li during PMS contraction and (ii) developed a radiative core that could halt any rapid Li depletion following a subsequent accretion or engulfment event. Observing stars in clusters offers the benefits of much more reliable estimates of age and mass than in field stars and a ready-made population of comparison siblings, born with similar initial chemical composition, from which deductions might be made about the timing and frequency of the engulfment of planetary material.

The observational material and search methodology are described in \S\ref{sourcedata}. Li-rich outliers are investigated and compared with other cluster members in \S\ref{results} and the results are discussed in the context of several hypotheses, including planetary engulfment, in \S\ref{discussion}. A summary of findings is given in \S\ref{summary}.

\section{Lithium-rich M-dwarfs observed in the Gaia-ESO survey}

\label{sourcedata}

\subsection{Source data}

The {\it Gaia}-ESO Spectroscopic (GES) survey \citep{Gilmore2022a, Randich2022a} provides the source data within which to search for Li-rich low-mass stars in clusters. GES obtained spectra for thousands of stars in 63 open clusters over a wide range of ages, using (pre-{\it Gaia}) photometry as the principle selection criteria for PMS and main sequence stars \citep{Bragaglia2022a}. \cite{Jackson2022a} combined GES spectroscopic parameters (principally $T_{\rm eff}$ and radial velocities) with {\it Gaia} astrometry to identify cluster members. The target selection and membership assessment were unbiased with respect to the Li content of the stars.

The equivalent width of the Li~{\sc i}~6708\AA\ resonance line (EWLi) is the principal Li abundance indicator. EWLi was measured for all GES cluster members in the work of \cite{Jeffries2023a}, who used these along with a set of adopted cluster ages to calibrate the {\sc eagles} model, which can estimate the age of individual stars or the common age of a group of stars, given their EWLi and $T_{\rm eff}$. 
Further refinements were presented by \cite*{Jackson2025a} who filtered out a small percentage of previously assigned members on the basis of anomalous parallaxes or {\it Gaia} photometry and by \cite*{Weaver2024a} who used an artificial neural network (ANN) to improve the {\sc eagles} model.

\begin{figure*}
    \centering
    \includegraphics[width=0.96\textwidth]{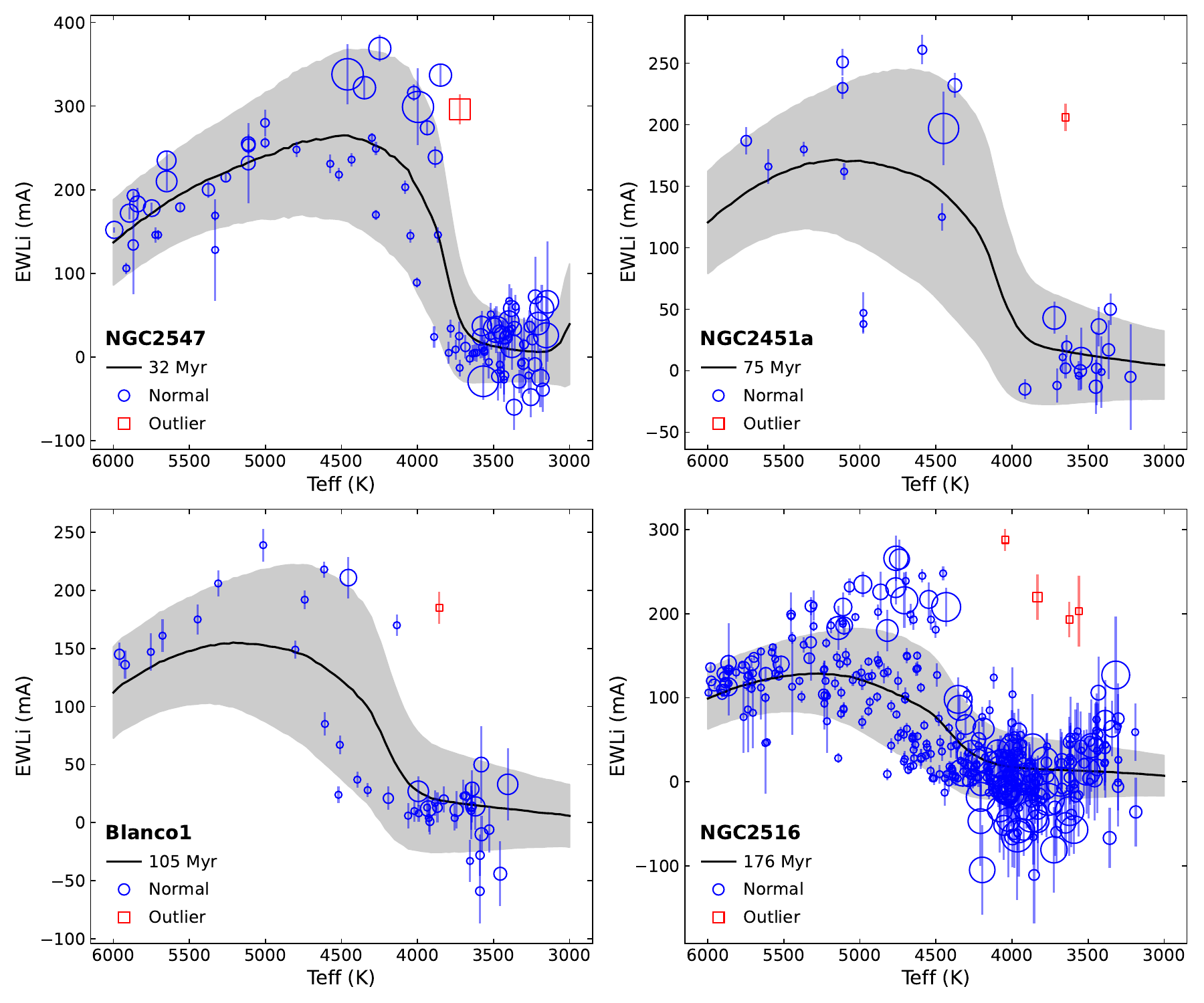}
\caption{The EWLi-$T_{\rm eff}$ data for the four clusters where Li-rich outliers are detected (identified by red squares). The symbol diameter scales as $d = d_{\rm min} + 3\log_{10}(v\sin i$/km\,s$^{-1}$), with $d=d_{\rm min}$ for $v\sin i \leq 10$ km\,s$^{-1}$. The solid curves show the most likely {\sc eagles} isochrones fitted to all the cluster data and the grey shaded regions represent the {\sc eagles} model estimate of the $1\sigma$ normal intrinsic dispersion of EWLi at that age.}
\label{liplots}
\end{figure*}

Here we use the catalogue of cluster members, EWLi and $T_{\rm eff}$ from \cite{Jackson2025a}, to identify stars that are anomalously Li-rich compared to cluster siblings of similar $T_{\rm eff}$. \cite{Jackson2025a} only presented data for 11 GES clusters with $30<$ age/Myr $<370$. To these we added IC~2391 and IC 2602, which Jackson et al. excluded because they had too few low-mass K-type targets for the analysis in that paper and NGC~6281 and NGC~6633, that are the only older clusters in GES where M-type cluster members with $T_{\rm eff}<4100$\,K were observed. 

The 15 clusters are summarised in Table~\ref{clustertable}. The ages adopted for the clusters are those described and discussed by \cite{Jackson2025a} and for the additional clusters we also adopt the ages that were used in training the {\sc eagles} model \citep{Jeffries2023a}. For each cluster we report $N_T$, the total number of EWLi measurements covering stars from $3000 \leq T_{\rm eff}/{\rm K} < 6500$, and $N_M$, the number of late-K/early-M dwarfs with $3500 < T_{\rm eff}/{\rm K} < 4100$.

\begin{table}
\centering
\caption{The clusters considered here. "Age" is the age adopted when calibrating the {\sc eagles} models \citep{Jeffries2023a}, whilst "Li age" is the most likely age determined by fitting the EWLi data for that cluster using {\sc eagles}.}
\begin{tabular}{
    l 
    S[table-format=4.0] 
    c 
    S[table-format=3.0] 
    S[table-format=3.0] 
    S[table-format=3.0]
}
\toprule
Cluster & \multicolumn{1}{c}{Age} & \multicolumn{1}{c}{Li Age} & \multicolumn{1}{c}{$N_T^{\dagger}$} & \multicolumn{1}{c}{$N_M^{\ddagger}$} & \multicolumn{1}{c}{$N_{\rm out}^{*}$} \\
        & \multicolumn{2}{c}{(Myr)} &      &       &                \\
\midrule
NGC 2547 & 35 & $32^{+2}_{-2}$ & 98 & 30 & 1\\
NGC 2451b   & 38 & $41^{+5}_{-5}$ & 57 & 15 & 0\\
IC 2391 & 42 & $62^{+45}_{-15}$ & 33 & 7 & 0\\
IC 2602 & 43 & $42^{+6}_{-4}$ & 54 & 12 & 0\\
NGC 2451a & 50 & $75^{+28}_{-13}$ & 29 & 10 & 1 \\
IC 4665   & 55 & $45^{+7}_{-7}$  & 31 & 8 & 0 \\
NGC 6405 & 63 & $77^{+17}_{-13}$ & 44 & 4 & 0 \\
Blanco 1 & 110 & $105^{+21}_{-20}$ & 47 & 27 & 1 \\
Pleiades & 120 & $76^{+9}_{-7}$ & 88 & 6 & 0 \\
M35      & 140 & $117^{+11}_{-7}$ & 211 & 0 & 0 \\
NGC 6709 & 166 & $158^{+24}_{-29}$ & 32 & 0 & 0 \\
NGC 2516 & 209 & $176^{+10}_{-8}$ & 386 & 140 & 4 \\
NGC 3532 & 367 & $437^{+20}_{-20}$ & 276 & 49 & 0\\
NGC 6281 & 375 & $324^{+61}_{-48}$ & 23 & 6 & 0\\
NGC 6633 & 622 & $1160^{+1690}_{-380}$ & 17 & 4 & 0 \\

\bottomrule
\multicolumn{6}{l}{$\dagger$ Total stars with $3000<T_{\rm eff}/{\rm K}<6500$.} \\
\multicolumn{6}{l}{$\ddagger$ Number of M-stars with $3500<T_{\rm eff}/{\rm K}<4100$.} \\
\multicolumn{6}{l}{$*$ The number of $>3.2\sigma$ outliers found (see \S\ref{s_outliers}).}
\end{tabular}
\label{clustertable}
\end{table}

\subsection{Selecting Li-rich candidates}
\label{s_outliers}

For each cluster, the ANN version of the {\sc eagles} model was used to establish a most-likely EWLi-$T_{\rm eff}$ isochrone along with a 1-sigma intrinsic dispersion \citep[the dispersion beyond that explained by observational uncertainties -- see][]{Weaver2024a}. The age of the most-likely isochrone is not necessarily the same as the training age adopted for that cluster, though it is similar in most cases. We did this to allow the possibility that there are cluster-to-cluster differences in Li-depletion that are not fully captured by the {\sc eagles} model. 

We then search for stars that are $>3.2\sigma$ above the most-likely isochrone for the cluster, where $\sigma$ was evaluated as the quadrature sum of the intrinsic dispersion and the observational uncertainty in EWLi for each star. The choice of $3.2\sigma$ as a threshold means that in a total sample of $N$ stars then $6.9\times 10^{-4}N$ would be expected to exceed that threshold in a normal distribution. Thus we would expect just 1 outlier in the full sample of clusters in Table~\ref{clustertable} ($\Sigma N_T =1426$) and only 0.2 outliers in the M-star range ($\Sigma N_M=318$), if they were normally distributed.

\begin{table*}
    \caption{The properties of the 7 Li-rich late-K/M-dwarfs.}
    \label{outliers}
\begin{tabular}{lcccc}
   \toprule
\midrule
$^1$GES  & 07581528-6059334 & 07584014-6107216 & 07591796-6045189 & 08024838-6047096 \\
\midrule
Cluster & NGC~2516 & NGC~2516 & NGC~2516 & NGC~2516 \\
Gaia  &  5290668915697538560 & 5290655412320662144 & 5290768284061176192 & 5290756460011285120\\
$^2 P_{\rm mem}$ & 0.9999 & 0.9999 & 0.9995 & 0.9998 \\
\\
$T_{\rm eff}$ (K)  & $3560\pm 100$ & $4045\pm 74$ & $3833\pm 109$ & $3622\pm 104$ \\
EWLi (m\AA) & $203\pm 42$ & $288 \pm 13$ & $220 \pm 27$ & $193\pm 21$ \\
$A$(Li)       &   $1.3\pm 0.4^{8}$          & $2.00 \pm 0.08$ & $1.69 \pm 0.16$ & $1.15 \pm 0.23$ \\
$A$(Li) (3D NLTE) & 1.3 & 1.87 & 1.61 & 1.18 \\
$[$Fe/H$]$      &             & $0.04 \pm 0.09$ & $-0.22 \pm 0.09$ & $-0.27 \pm 0.09$ \\
$v\sin i$ (km\,s$^{-1}$) & 8 & 4 & 13 & 7 \\
$^3$Period (d) & 3.319 & 9.314 & & 10.406 \\

\\
$G$  & 17.33 & 16.33 & 16.69 & 17.31\\
$G_{BP}-G_{RP}$  & 2.38 & 1.97 & 2.35 & 2.34 \\
$G-G_{RP}$ & 1.13 & 0.99 & 1.12 & 1.10 \\
RUWE  & 1.03 & 1.08 & 1.40 & 0.99 \\ 
$^4 T_{\rm SED}$ (K) & $3600 \pm 50$ & $3900\pm 50$ & $3600\pm 50$ & $3600 \pm 50$ \\ 
\\
$^5$Z-score  & 3.57 & 6.27 & 4.51 & 4.54 \\  
$^6$Age (Myr)  & 209 & 209 & 209 & 209 \\ 
Mass ($M_\odot$) & $0.49\pm 0.03$ & $0.64 \pm 0.02$ & $0.59 \pm 0.02$ & $0.52 \pm 0.02$ \\
$^7$Li-age (Myr) & $22^{+46}_{-5}$ & $30^{+15}_{-9}$ & $30^{+7}_{-8}$ & $23^{+4}_{-7}$ \\
\\
\midrule
$^1$GES & 08101842-4902145 & 07460836-3753518 & 00022426-3009090 & \\
\midrule
Cluster & NGC~2547 & NGC~2451a & Blanco~1 & \\
Gaia  & 5514374005300862336 & 5538809689156020992 & 2320793691588062464 & \\
$^2 P_{\rm mem}$ & 0.9871 & 0.9997 & 0.9999 & \\
\\
$T_{\rm eff}$ (K) & $3721 \pm 55$ & $3648\pm 54$ & $3857\pm 59$ &  \\
EWLi (m\AA) & $296\pm 18$ & $206\pm 11$ & $185\pm 14$ &  \\
$A$(Li) & $2.07\pm 0.02$  & $1.42 \pm 0.07$  & $1.52 \pm 0.09$  & \\
$A$(Li) (3D NLTE)  & 1.94 & 1.39 & 1.47 & \\
$[$Fe/H$]$ &              &             &  $-0.17 \pm 0.08$       & \\
$v\sin i$ (km\,s$^{-1}$) & 46 & 8 & 8 & \\
$^3$Period (d) & 1.82 & & 5.882 & \\
\\
$G$ & 15.62 & 14.98 & 15.30 & \\
$G_{BP}-G_{RP}$ & 2.21 & 2.27 & 2.01 &  \\
$G-G_{RP}$ & 1.07 & 1.09 & 1.00 & \\
RUWE & 1.04 & 1.07 & 1.07 &  \\ 
$^4 T_{\rm SED}$ (K) & $3600\pm 50$ & $3600\pm 50$ & $3800\pm 50$ & \\
\\
$^5$Z-score & 3.40 & 4.27 & 3.49 &  \\  
$^6$Age (Myr) & 35 & 50 & 110 & \\ 
Mass ($M_{\odot}$) & $0.65\pm 0.01$ & $0.60\pm 0.01$ & $0.57 \pm 0.02$ & \\
$^7$Li-age (Myr) & $22^{+4}_{-7}$ & $24^{+3}_{-5}$ & $34^{+10}_{-8}$ & \\

\bottomrule

\multicolumn{5}{l}{$1$ Identifier in the GES survey.}\\
\multicolumn{5}{l}{$2$ 3D kinematic membership probabilities from \protect\cite{Jackson2022a}.}\\
\multicolumn{5}{l}{$3$ For sources of rotation periods see \S\ref{S_Discuss1}.}\\
\multicolumn{5}{l}{$4$ Temperature determined from the optical/near-IR spectral energy distribution.}\\
\multicolumn{5}{l}{$5$ The number of sigma by which EWLi exceeds a predicted value (see \S\ref{s_outliers}).}\\
\multicolumn{5}{l}{$6$ The adopted fiducial age of the cluster in \protect\cite{Jackson2025a}.}\\
\multicolumn{5}{l}{$7$ The age inferred from the star's EWLi and $T_{\rm eff}$ (see \S\ref{S_Discuss1}).}\\
\multicolumn{5}{l}{$8$ Li abundance is our estimate based on EWLi and $T_{\rm eff}$.}
\end{tabular}

\end{table*}

\section{Results}

\label{results}

\begin{figure}
    \centering
    \includegraphics[width=0.44\textwidth]{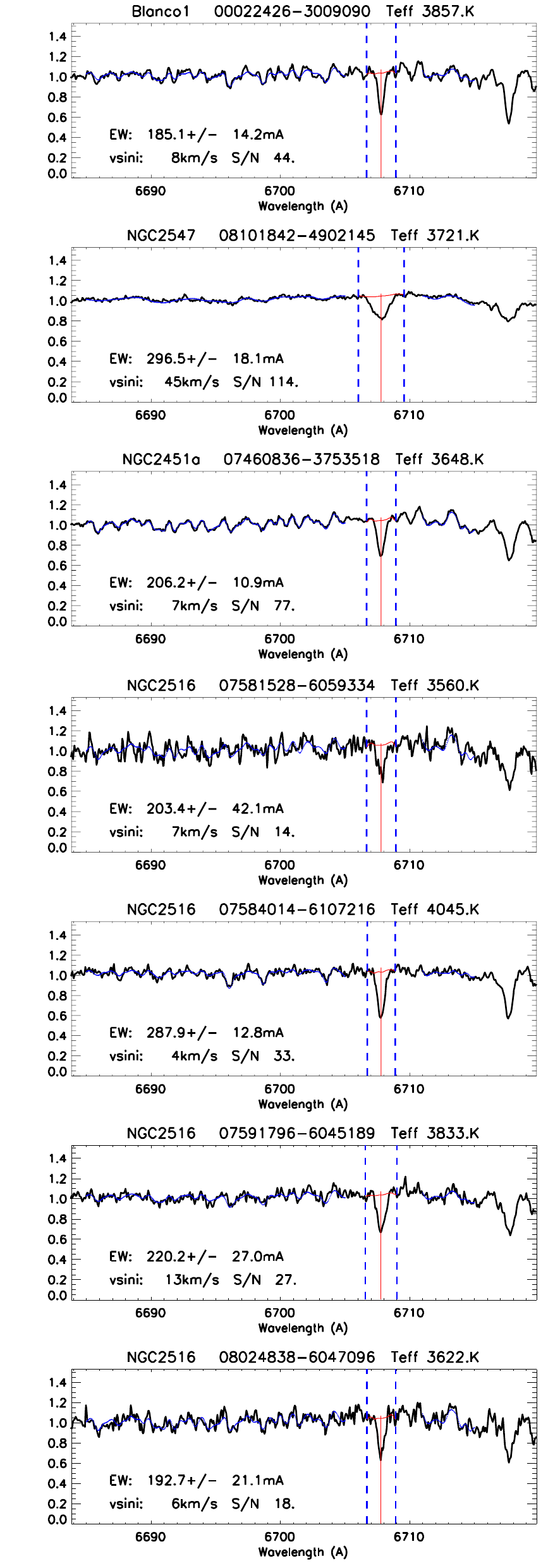}
\caption{Normalised spectra around the Li~{\sc i}~6708\AA\ line for the 7 Li-rich outliers (black lines). The blue line in each plot shows the scaled and broadened template spectrum used to fix the continuum level. The red lines show the template spectra immediately around the Li line and between the equivalent width integration limits (shown as vertical dashed lines).}
\label{specplot}
\end{figure}  

\begin{figure}
    \centering
    \includegraphics[width=0.44\textwidth]{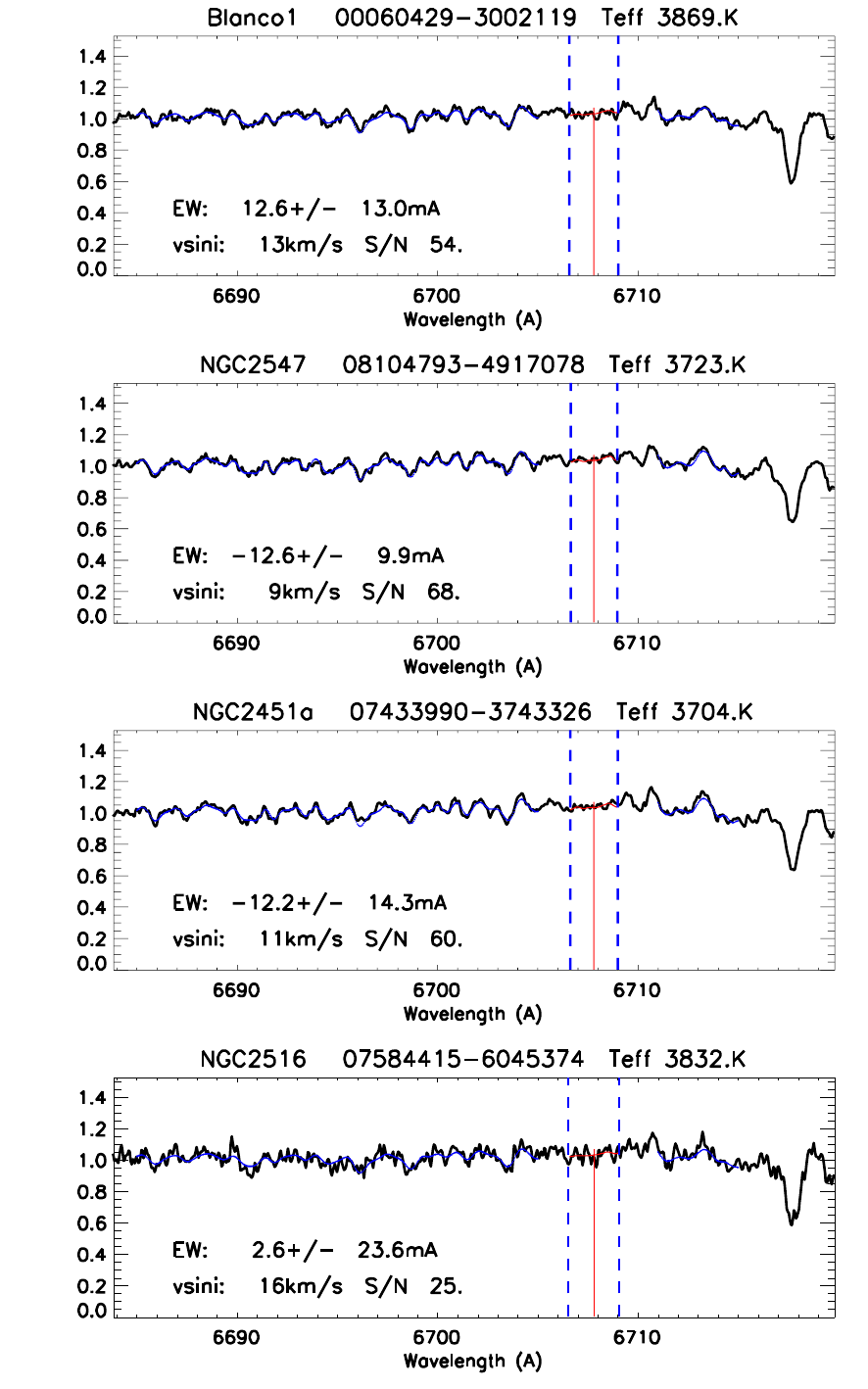}
    \caption{Similar to Fig.~\ref{specplot} but these are representative spectra of the majority Li-poor stars in these clusters, with similar $T_{\rm eff}$ to the outliers.}
    \label{specplot_normal}
\end{figure}

\begin{figure*}
    \centering
    \includegraphics[width=0.95\textwidth]{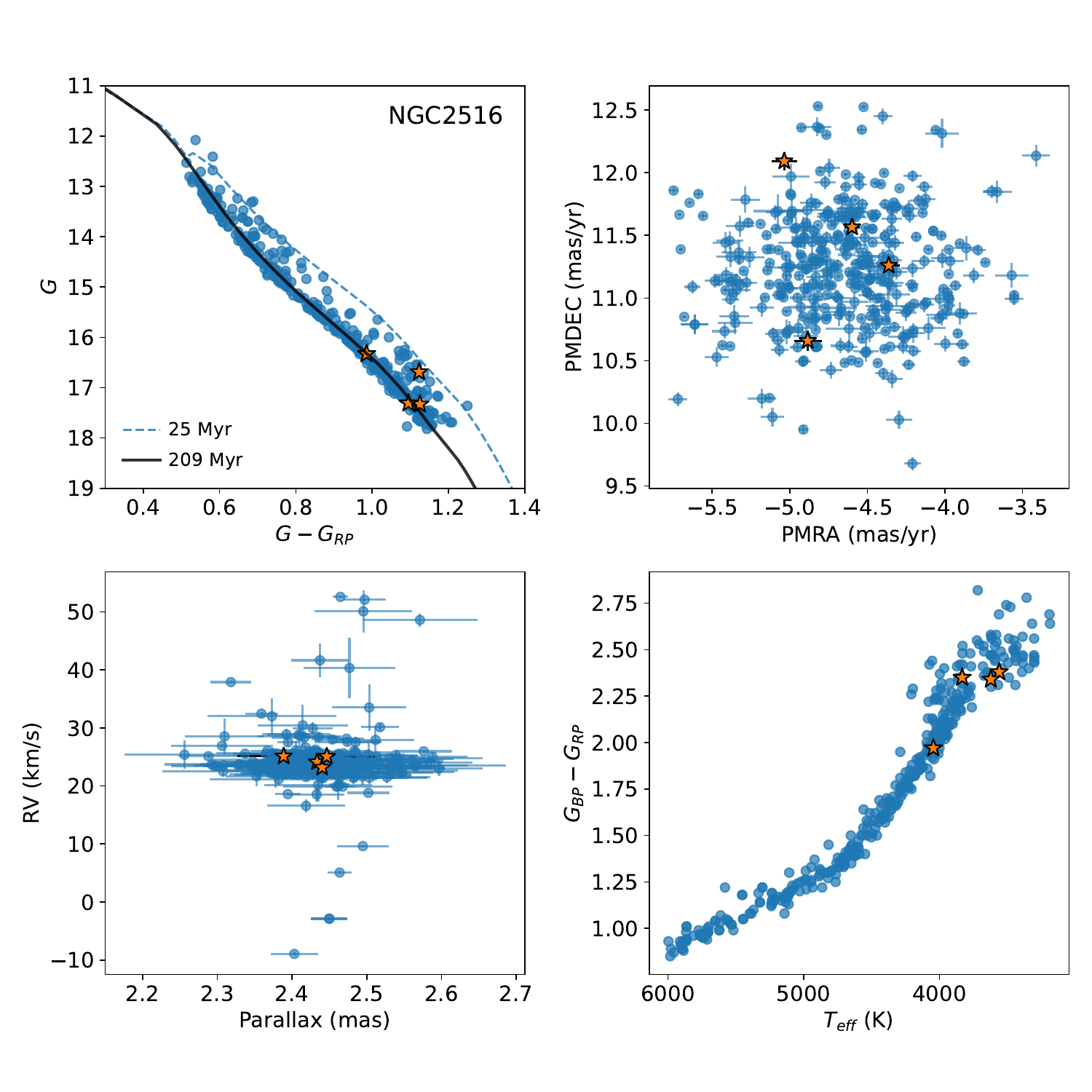}
\caption{The position of the 4 Li-rich outliers in NGC~2516 compared with other cluster members in several diagnostic diagrams. The top-left plot shows isochrones from the {\sc spots} models of \protect\cite{Somers2020a} at the nominal age of the cluster and at 25\,Myr (i.e. where a star of age 25\,Myr but at the same distance as NGC~2516 would appear)}
\label{NGC2516plot}
\end{figure*}

\subsection{The Li-rich outliers}

\label{results1}

We find 7 anomalously Li-rich outliers beyond the $3.2\sigma$ threshold, that are members of 4 clusters -- NGC~2547 (1 outlier), NGC~2451a (1 outlier), Blanco~1 (1 outlier) and NGC~2516 (4 outliers) -- see Fig.~\ref{liplots}.  The outliers and their relevant properties are listed in Table~\ref{outliers}. The spectra of these objects are shown in Fig.~\ref{specplot}, which demonstrates the excellent quality of the data and leaves no doubt as to the reliability of the Li~{\sc i} signature. For comparison, Fig.~\ref{specplot_normal} shows spectra, that are representative examples of the majority of stars in these clusters with similar $T_{\rm eff}$, that have no sign of the Li\,{\sc i} line at all. 

As a final precaution against instrumental or data reduction artifacts, we checked that the strong Li signatures in the spectra could not be due to contamination by adjacent Li-rich stars in the fibre setups in which they were configured. We confirmed that none of the Li-rich outliers had brighter stars either side of them in the recorded CCD images of their spectra and as such should have suffered no contamination of their spectra during extraction.

All of the outliers are in the narrow temperature range $3550<T_{\rm eff}/{\rm K}<4050$ (and narrower when considering the spectral energy distributions -- see \S\ref{S_Discuss1} and Appendix~\ref{AppSED}). The overall frequency of Li-rich outliers in this range is therefore $7/318 \simeq 2.2^{+0.6}_{-0.5}$ percent (68 per cent confidence interval). The expected number of outliers in this $T_{\rm eff}$ range, for a normally distributed dispersion, is 0.2, so the probability of seeing 7 or more by chance is $\sim 10^{-9}$. This suggests that there is a non-Gaussian tail of objects  with anomalously high EWLi in this $T_{\rm eff}$ range, confirming the visual impression given by Fig.~\ref{liplots}.

To analyse the properties of the outliers further, their locations are compared with all their siblings in a $G$ versus $G_-G_{RP}$ colour-magnitude diagram (CMD), a proper motion diagram, a radial velocity versus parallax diagram and a $G_{BP}-G_{RP}$ versus $T_{\rm eff}$ diagram for each of their host clusters. The purpose is to see whether these Li-rich outliers are unusual in any other way. An example is shown in Fig.~\ref{NGC2516plot} for the 4 outliers identified in NGC~2516. The plots for the other three clusters are in Appendix~\ref{App1} (Figs.~\ref{NGC2547plot}, \ref{NGC2451aplot} and \ref{Blanco1plot}).

In the CMD we also compare the positions of stars with isochrones from the {\sc spots} models \citep{Somers2020a} at the nominal cluster ages listed in Table~\ref{clustertable}, using the intrinsic distance modulus and reddening listed in \cite{Jackson2022a} and assuming an average $A_G \simeq 2.74 E(B-V)$ and $E(G-G_{RP})\simeq 0.7E(B-V)$ \citep{Casagrande2018a}. The models with a spot coverage fraction of 30 per cent were chosen. These have been found to represent the CMDs of young clusters quite well \citep[e.g.,][]{Binks2022a} and match spectroscopic estimates of spot coverage in the low-mass stars of several young clusters \citep{Cao2025}. These models provide a reasonable representation of the single star sequences in the cluster CMDs here and the mass of the outliers is estimated from their $T_{\rm eff}$ using these isochrones.

Table~\ref{outliers} also contains the Li abundance, $A$(Li) $= 12 + \log_{10} (N({\rm Li})/N({\rm H}))$, derived from an LTE analysis \citep{Franciosini2022a} and the [Fe/H] \citep{Hourihane2023a}, as reported in the GES DR4 parameter catalogue. For one target in NGC~2516 no $A$(Li) value was given in GES DR4 and we estimated it by comparison with stars of similar EWLi and $T_{\rm eff}$ in the GES catalogue. Approximate 3D NLTE abundances were estimated using the corrections of \citet{Wang2021a}, with corrections for stars with $T_{\rm eff}<4000$\,K (outside the grid of corrections) fixed at their values for $T_{\rm eff}=4000$\,K.

\subsection{Are the outliers cluster members?}

\label{S_Discuss1}

Figure~\ref{NGC2516plot} shows that the 4 Li-rich outliers in NGC~2516 appear completely "normal" in terms of their parallax and kinematic properties, from which their very high probability of cluster membership ($>0.999$ in each case) arises. The CMD shows that one outlier is probably an unresolved binary, more luminous by 0.6 mag than the single star main sequence. This star (J07591796-6045189) has a radial velocity consistent with the cluster mean, suggesting it probably isn't a close binary and its moderately large {\it Gaia} RUWE parameter of 1.40, is also consistent with a wide binary scenario \citep{CastroGinard2024}. The other three outliers in NGC~2516 lie on the single star main sequence of the cluster and have RUWE values $\sim 1.0$, consistent with single star status.

Looking at Fig.~\ref{liplots} (lower-right) it appears the 4 outliers in NGC~2516 form a quite separate population from the very scattered K-stars at $4300< T_{\rm eff}/{\rm K} < 5400$. The latter exhibit the well-known correlation between rapid rotation and higher EWLi that has been identified in the K-type stars of a number of young clusters \citep{Bouvier2018a, Jeffries2021a, Jackson2025a}. Perhaps some of the Li-rich stars in this warmer $T_{\rm eff}$ range with low $v \sin i$ are also anomalous, but identifying them is difficult because of the large intrinsic scatter. In contrast, the 4 cooler Li-rich stars we have identified here stand as very clear outliers against a large Li-depleted background population with smaller intrinsic scatter. None of them appear to be rapid rotators \citep[judged by $v \sin i$, but confirmed with slow rotation periods found from spot modulation for 3 of them --][and see Table~\ref{outliers}]{Irwin2007a, Fritzewski2020a} and there are many faster-rotating Li-depleted stars at similar $T_{\rm eff}$.

The outliers could perhaps be explained if their $T_{\rm eff}$ has been drastically underestimated by at least 500\,K. The intrinsic dispersion estimated by the {\sc eagles} model already factors in the typical $\sim 100$\,K uncertainties of the GES spectroscopic $T_{\rm eff}$ measurements and Fig.~\ref{NGC2516plot} (lower-right panel) shows that these 4 stars have spectroscopic $T_{\rm eff}$ estimates that are consistent with their $G_{BP}-G_{RP}$ colours. As a further consistency check, we modelled the optical and near-IR spectral energy distributions (see Appendix~\ref{AppSED}), finding $3600 \leq T_{\rm SED}/{\rm K} \leq 3900$ (see Table~\ref{outliers}), in reasonable agreement ($< 2 \sigma$) with the GES spectroscopic estimates in all cases.

A possibility to consider is that these Li-rich stars could just be young field stars in the direction of NGC~2516. In terms of sky position they lie well within the distribution of other cluster members. They also have parallaxes that put them within a few pc of the cluster centre. Thus they are spatially coincident with the cluster. Their individual EWLi and $T_{\rm eff}$ measurements can be used in the {\sc eagles} model to give a posterior probability distribution for their age, assuming a flat prior probability. This returns ages (see Table~\ref{outliers}) of 22--30\,Myr that are of course significantly younger than the cluster. However, if we look at the CMD in Fig.~\ref{NGC2516plot}, where a 25\,Myr {\sc spots} isochrone is plotted, stars of age 22--30\,Myr at the same distance as NGC~2516 should be about 1 mag brighter than the cluster's single star sequence and brighter even than equal-mass binaries in NGC~2516. We conclude that the Li-rich stars cannot be as young as their EWLi suggests.

Similar arguments apply to the single outliers in NGC~2451a and Blanco 1 (see Figs.~\ref{NGC2451aplot} and \ref{Blanco1plot}). These sit on the single star sequence of their respective clusters, show no indication of binarity either in their radial velocities or RUWE, and their spectroscopic $T_{\rm eff}$ estimates agree with those from SED modelling. They are slow rotators \citep[confirmed by a 5.9 d rotation period for the Blanco 1 outlier,][]{Gillen2020a} and are significantly cooler than the K-type stars with lithium in those clusters (Fig.~\ref{liplots}). The outlier in NGC~2547 is more debatable (see Fig.~\ref{NGC2547plot}). It is separated from warmer, Li-rich siblings by only $\sim 100$\,K (or perhaps 200\,K from SED modelling) . Its status as an outlier is due to the sharp drop in the mean EWLi-$T_{\rm eff}$ relation at only slightly higher temperatures and is vulnerable to any small $T_{\rm eff}$ underestimate. Further, it appears to be a photometric binary and its radial velocity is displaced from the cluster mean by $\sim 8$ km\,s$^{-1}$, indicating a possible short orbital period. It also has a high $v\sin i$ and a moderately short rotation period of 1.8\,d \citep{Irwin2008b}, a characteristic it shares with the only slightly warmer Li-rich stars. Its position in the CMD could also be compatible with it being a slightly younger interloper.

In summary, the analyses here show that all 7 Li-rich outliers are indistinguishable from cluster members in all other respects -- colours and magnitudes, kinematics and parallax. Of these, 6 have a position in the CMD that is incompatible with the youth implied by their Li abundance and $T_{\rm eff}$, appear distinct from their siblings in the EWLi vs $T_{\rm eff}$ diagram, and show no signs of anomalous reddening or $T_{\rm eff}$ underestimates.

\section{Discussion}

\label{discussion}

\begin{figure}
    \centering
    \includegraphics[width=0.48\textwidth]{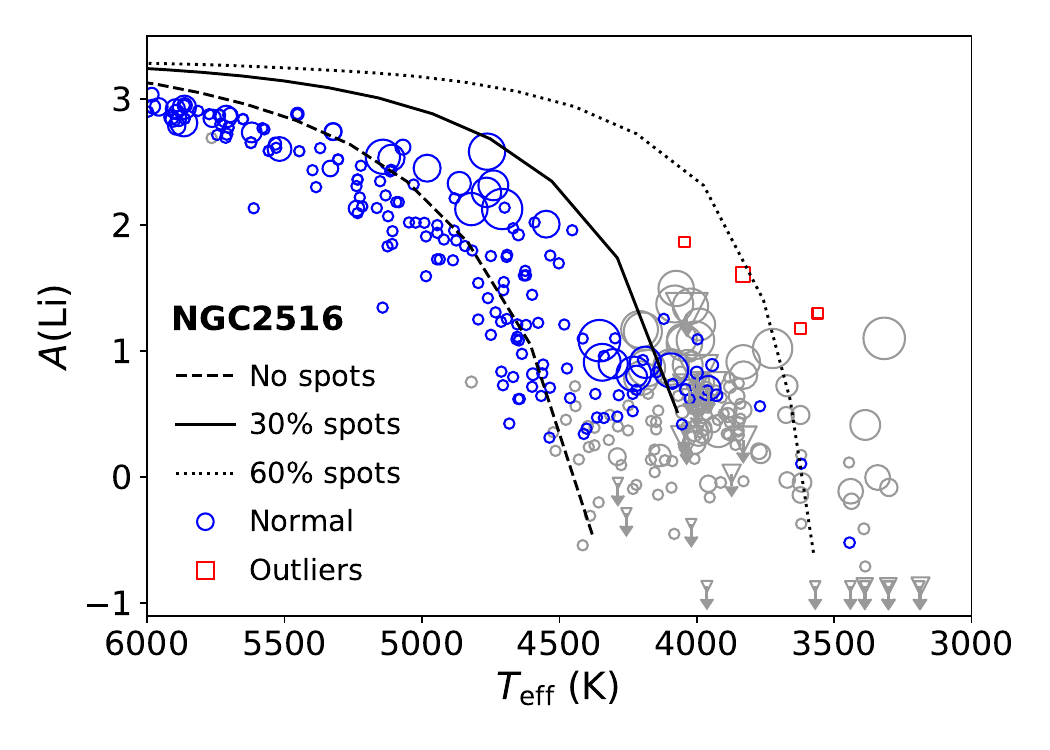}
\caption{The position of the 4 Li-rich outliers in NGC~2516 in an $A$(Li) versus $T_{\rm eff}$ diagram. 
The observational data are from the GES DR5 abundance catalogue with downward arrows indicating abundance upper limits. The size of the points is scaled with $v \sin i$ as in Fig.~\ref{liplots}. The points shaded in grey are those which have an EWLi \protect\citep[in][]{Jeffries2023a} smaller than twice their error bar and which arguably should all be treated as upper limits (see \S\ref{inhibit}). Isochrones of Li depletion from the {\sc spots} models of \protect\cite{Somers2020a}, at the nominal age of the cluster, are shown for spot areal coverage fractions of 0, 30 and 60 per cent.}
\label{NGC2516Aliplot}
\end{figure}  

At least 6 of the outliers appear  to be genuine members of their respective clusters with anomalously strong Li line strengths. Here we consider three scenarios that might explain them.

\begin{itemize}
    \item That their magnetic activity or rotation has been unusually effective in inhibiting the normal PMS Li depletion experienced by their siblings.

    \item That peculiarities in their initial conditions or formation process has left these stars with more Li than expected.

    \item That these stars have been ``polluted" by the accretion of Li-enriched planetary material.
    
\end{itemize}

\subsection{An inhibited Li depletion scenario}

\label{inhibit}

Dynamo-generated magnetic activity or just fast rotation could inhibit processes that lead to the mixing and burning of Li during the PMS phase. A range of activity and rotation among coeval stars has been put forward as a means of explaining the dispersion in Li abundance and observed correlation between fast rotation (and greater magnetic activity) and less Li depletion seen among late-G and K-dwarfs in ZAMS and PMS clusters \citep[e.g.,][and references therein]{Bouvier2018a, Jeffries2021a, Jackson2025a}. Possible mechanisms, supported by modelling, are the inhibition of convective flux by magnetic fields in the interior \citep{Ventura1998a, Feiden2016a} or due to starspots at the surface \citep{Somers2014a, Somers2015b}; or a reduction in the efficiencies of convection or of overshooting into the radiative core, due to fast rotation \citep{Baraffe2017a, Constantino2021a}.

As an example of how this might work to explain the Li-rich outliers, Fig.~\ref{NGC2516Aliplot} shows NLTE-corrected Li abundances for NGC~2516 from the GES DR4 catalogue compared with predictions of the {\sc spots} models at 209\,Myr, for spot coverage fractions of 0, 30 and 60 per cent. The outliers are marked with red squares and the symbols are scaled in size according to the measured $v \sin i$. Similar plots for the other clusters are given in Appendix~\ref{AppALi}. 

As expected, the outliers have the highest Li abundances for stars of their $T_{\rm eff}$; their outstanding nature is not so obvious when considering a (logarithmic) abundance plot. However, we note that many of the quoted Li abundances for M-dwarfs in the GES catalogue are based on EWLi values that are small or noisy. In Fig.~\ref{NGC2516Aliplot}, the de-emphasised grey points (both claimed detections and upper limits) are those for which EWLi is less than twice its uncertainty (and in most cases, compatible with zero within its uncertainty). These should probably all be considered upper limits (or unreliable).

\begin{figure}
    \centering
    \includegraphics[width=0.48\textwidth]{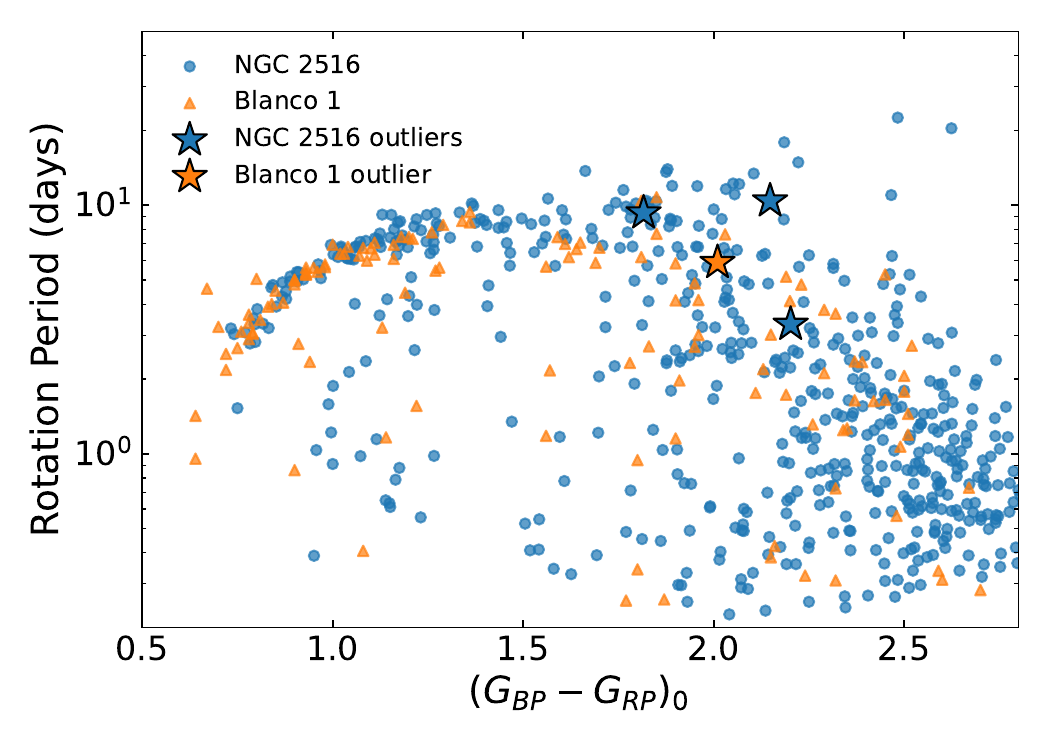}
\caption{The outliers with rotation periods compared with other cluster members in a rotation period versus intrinsic $(G_{\rm BP}-G_{\rm RP})_0$ diagram  \protect\citep[data from][]{Fritzewski2020a, Gillen2020a}. }
\label{rotplot}
\end{figure}  

The {\sc spots} models assume spot coverage is consistent throughout the PMS evolution and that Li depletion occurs solely due to convective mixing and burning. The lower envelope of the NGC~2516 distribution is actually lower than the zero spots isochrone at $T_{\rm eff}>5000$\,K. That is not the case in the younger clusters (see Appendix~\ref{AppALi}) and it appears that some extra Li depletion has occurred among the warmer NGC~2516 stars. The faster rotators (as judged by $v \sin i$) tend to define the upper envelope of the distribution. It is worth noting that the isochrones plotted at 110\,Myr for Blanco 1 (Fig.~\ref{AppALiplot}) appear almost identical, because convectively-driven PMS Li depletion has essentially ended for any stars with detectable Li once the ZAMS is reached at $\sim$50\,Myr.

To explain the outliers using these models would require spot coverage fractions of 50-70 per cent and, for NGC~2516, Blanco~1 and NGC~2451a, such a model would predict significantly less Li depletion than is seen for the upper envelope defined by warmer stars, where spot fractions of $\leq 30$ per cent would seem a better match. Of course it might be possible that lower mass stars do have larger spot coverage or stronger magnetic fields. Spectroscopically estimated spot coverage in ZAMS K- and M-dwarfs of the Pleiades \citep{Cao2022a} appears to saturate at $\leq 25$ per cent for the fastest rotators, but similar techniques reveal spot coverages can be much higher in younger K/M-type PMS stars at $\leq 10$ Myr \citep{Cao2025, PerezPaolino2025a} and perhaps could be maintained during the epoch of PMS Li depletion at 10--50\,Myr. In any case, these Li depletion predictions are extremely sensitive to the ingredients of physical models that determine the position and temperature of the CZ base -- for example the detailed chemical composition, treatments of opacity, convection and the simplifications needed to deal with magnetic fields and starspots in 1D models \citep[e.g.,][]{Piau2002a, Feiden2013a, Somers2020a}, so the plausibility of the required high spot coverage is not necessarily a problem. The problem is to explain why these 6 outliers should possess, or have possessed, a much larger spot coverage or larger magnetic fields than their siblings.

As discussed in \S\ref{S_Discuss1}, none of the 4 outliers in NGC~2516 or those in Blanco~1 and NGC2451a have high $v\sin i$. Rotation periods (listed in Table~\ref{outliers}) have been obtained from spot-induced light-curve modulation by \cite{Irwin2007a} and \cite{Fritzewski2020a} for 3 of the NGC~2516 outliers, and by \cite{Gillen2020a} for the Blanco~1 outlier. The distribution of rotation periods versus intrinsic colour in the NGC~2516 and Blanco~1 clusters, taken from the same sources, is shown in Fig.~\ref{rotplot}. This confirms (for those stars where we have periods) that the Li-rich outliers are slow rotators, in the upper quartile of rotation periods for stars of their colour and age. 

If the inhibition of Li depletion is connected with rapid rotation or magnetic activity and spots it would seem most likely to be more effective in the fastest rotators. The slow rotation of these outliers at the ZAMS is a strong argument against any such explanation. Although Li depletion takes place at earlier ages on the PMS, it also seems unlikely that these stars could have been amongst the fastest rotators then but among the slowest now.

\subsection{Peculiar initial conditions}

\label{initial}

The slow rotation of the outliers might implicate the earlier presence of a long-lived accretion disc. Slow rotation could be linked to a coupling between such a disc and the PMS star's magnetosphere, acting as a ``rotostat" to prevent spin up during PMS contraction \citep{Koenigl1991a, Armitage1996a}. Models show that longer disc-locking timescales lead to slower rotation at the ZAMS \citep{Gallet2015a, Vasconcelos2015a}. This idea is supported by the general observation that the fastest rotators in young PMS clusters tend to be those that are free of accretion signatures \citep{Herbst2002a, Cieza2007a}.

If the primordial accretion discs of these outliers were peculiarly long-lived, this might explain both the slow rotation and provide a late boost of pristine, comparatively Li-rich material after the main phase of PMS depletion has ended. However, this seems unlikely. In stars of the masses of the outliers, material accreted prior to $\sim 30$\,Myr would be rapidly depleted (see \S\ref{engulfment}). To enrich a star that would otherwise have lost all its Li and explain a final Li abundance of $\geq 10^{-2}$ of the initial abundance would require at least 1 per cent of the stellar mass to be accreted after that - implying steady accretion rates of $\dot{M}\sim 10^{-9}\,M_{\odot}$/yr over $\sim 10$\,Myr.
Such rates are not exceptional  in $0.5M_\odot$ stars at ages of 1-3\,Myr \citep{Manara2012a}, but surveys using accretion diagnostics that are sensitive to any $\dot{M} > 10^{-11}\,M_\odot/$yr, suggest a disc decay timescale of a few Myr and that only 2 per cent of stars at 10\,Myr are accreting even at this low threshold and none at older ages \citep{Fedele2010}.

Another possibility is that much earlier, non-steady or episodic, accretion could alter the internal structure of low-mass stars \citep{Baraffe2009a}. Initially motivated to explain the large fraction of embedded class I stars with low luminosity, it was found that, depending critically on the detailed accretion history and the fraction of the accretion energy radiated away by the protostar, episodic accretion might also explain the large luminosity spreads seen in the HR diagrams of very young PMS clusters and result in changes in core temperatures that could either hasten or delay Li depletion \citep{Baraffe2010a, Baraffe2017b}. In some scenarios the early growth of a radiative core can be triggered, lowering the temperature of the CZ base and inhibiting Li depletion in the envelope
compared with non-accreting models.

The problem with the episodic accretion idea in explaining Li-rich outliers at the ZAMS is that although Li-depletion is delayed it is not halted. Whilst some objects might still appear Li-rich at $10-30$\,Myr according to the models of \cite{Baraffe2017b}, the Li depletion rate in the lower mass stars is such that only those with $T_{\rm eff}>4000$\,K and $M>0.6M_\odot$ could avoid total depletion by 50\,Myr, which is not very different from the case of non-accreting models.

A final possibility that can be discarded, is that the outliers were born from material that was already anomalously Li-rich by at least an order of magnitude compared to material which forms otherwise similar stars in the same compact cluster. This scenario appears to be ruled out by the lack of any extremely Li-rich stars, with $A$(Li)\,$>4.3$ in younger clusters.

\subsection{A planetary engulfment scenario}
\label{engulfment}

\begin{figure}
    \centering
    \includegraphics[width=0.48\textwidth]{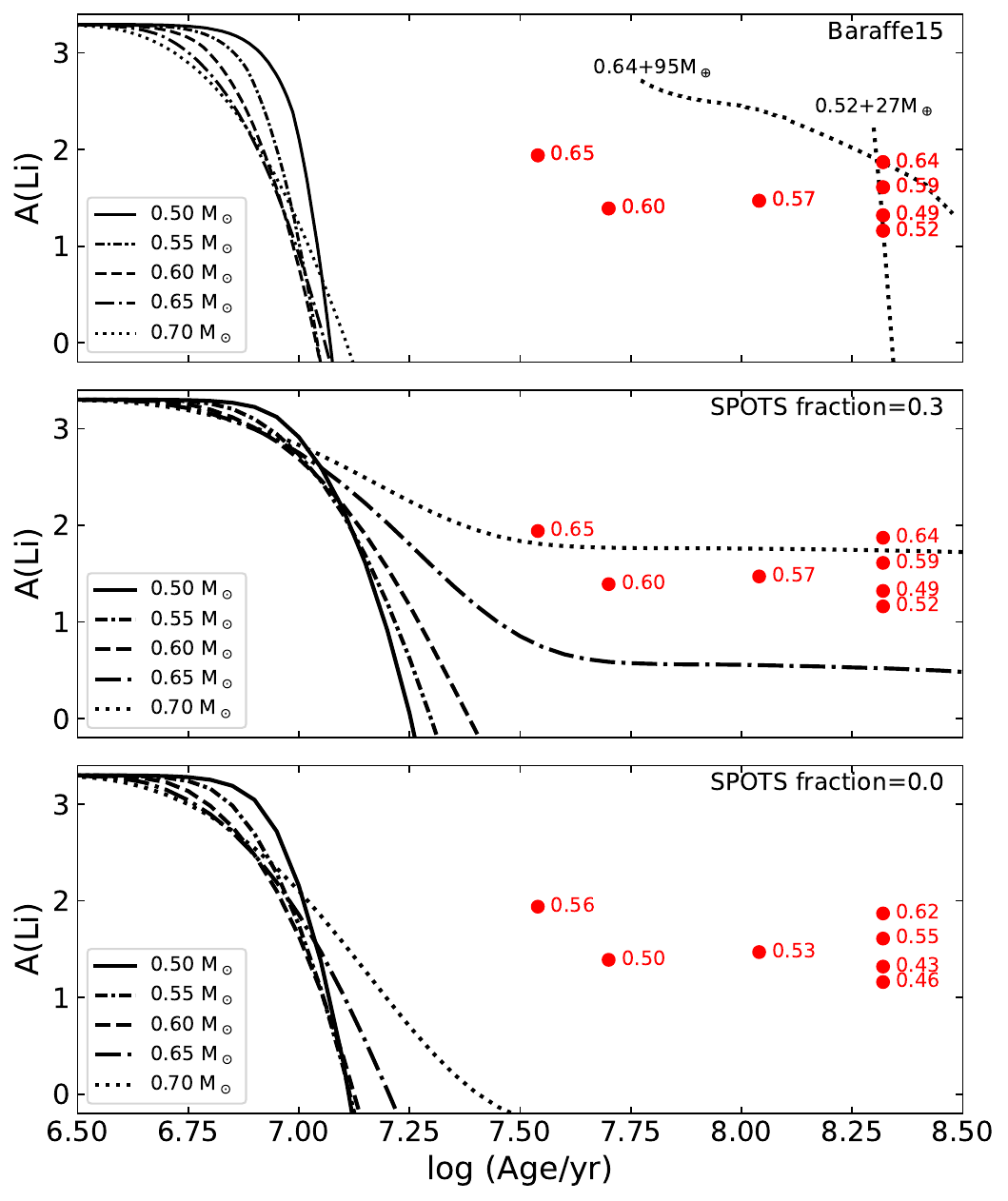}
\caption{Evolutionary tracks showing $A$(Li) as a function of time for several masses. The red points in each plot are the observed Li-rich stars, labelled with their masses (see Table~\ref{masses}). Top panel: Models from \protect\cite{Baraffe2015a}. The dotted lines with labels show examples of tracks featuring a planet engulfment at the time the track starts, with mass $10M_{\rm inst}$ (see Table~\ref{masses}), that then pass through the observed $A$(Li) and age of two of the Li-rich stars in NGC~2516. Middle panel: Tracks from the {\sc spots} models of \protect\cite{Somers2020a}, with a spot coverage of 30 per cent. Bottom panel: {\sc spots} models with zero spot coverage. The outliers are labelled with the (lower) masses that would be inferred (from their measured $T_{\rm eff}$) using these models.}
\label{lievolplot}
\end{figure}  

It is possible we have observational confirmation of the planetary engulfment scenario predicted by \cite{Sevilla2022a} -- the late accretion of planetary material by a star close to the ZAMS could result in an observable Li enhancement, that might survive for a significant duration if the accreting star is massive enough to have developed an extensive radiative core. The Li-rich outliers we find in Blanco~1 and NGC~2516, which should have reached the ZAMS, are consistent with Sevilla et al.'s scenario that a ZAMS accretion event of $10\,M_\oplus$ of Earth-like material results in an observable Li signature lasting 10-200\,Myr.

To provide our own quantitative assessment, the models described in \citet{Baraffe2015a} were used to estimate a baseline expected Li abundance at the age of the relevant clusters, starting from a uniform initial $A$(Li)\,$=3.3$ -- a reasonable value for solar-metallicity clusters near the Sun \citep{Randich2020a}. For the ages and masses of the 6 genuine outliers in Table~\ref{outliers} we confirmed that all their initial Li should have been destroyed - see the top panel of Fig.~\ref{lievolplot}.

We then calculated how much Li would have to be added instantaneously to the CZs of these stars (in the same models), at their cluster ages, in order to produce the observed $A$(Li). This can be turned into an estimate for an equivalent mass of Earth-like ``rocky" material by assuming an Li mass fraction of $1.1\times 10^{-6}$ for the bulk Earth \citep{McDonough1995a, McDonough2025a}.  The results are given in Table~\ref{masses}, listed under $M_{\rm inst}$ (in Earth masses) and are in the range 3-10$\,M_\oplus$.

There is some uncertainty in these estimates associated with the assumed composition of the accreted material, choice of evolutionary model and with the assumption we have made about 30 per cent spot coverage to estimate the masses. For example, $M_{\rm inst}$ could be about 30-40 per cent lower for material with the comparatively Li-enriched composition of Solar System meteorites \citep{Lodders2021a}. On the other hand, accretion of ISM-like or giant exoplanet material would require $M_{\rm inst}$ values $\sim 100$ times larger (1-3$\,M_{\rm Jupiter}$). At the same $T_{\rm eff}$, unspotted stars would have systematically lower inferred masses by $\sim 0.02-0.1\,M_\odot$ (see labels in the bottom panel of Fig.~\ref{lievolplot}), whereas more heavily spotted stars, or stars with spots that are warmer than the 80 per cent of the unspotted photospheric temperature assumed in the {\sc spots} models \citep[for which there is some evidence in M-dwarfs,][]{Berdyugina2005a}, could be more massive. However, this level of uncertainty in the stellar mass has only a small impact (15--30 per cent) on $M_{\rm inst}$, because a less massive star has a larger fraction of its mass in the CZ and vice-versa, and the CZ mass is little changed. The slightly larger required $M_{\rm inst}$ values for stellar masses reduced by $0.05\,M_\odot$ (shown in brackets) are also given in Table~\ref{masses} for comparison.

Following on from this, we can then ask the questions: how long ago could the engulfment have taken place and still produce the observed Li enhancement now and how long will the observed Li enhancement last? The answers to these questions are important in assessing the timing of, and how frequently, any engulfment of planetary material must occur within the population of low-mass stars. The answers are mass- and age-dependent since the rate of Li-depletion decreases with increasing mass at any given age, and decreases with increasing age at a given mass as the CZ base pushes outwards and cools. The answers do not depend on the current value of the Li enhancement, since the star's structure is not dependent on its Li content and to first order will not depend on how or how quickly a relatively small amount of a few Earth masses is accreted (engulfment of a Jupiter would be a different scenario).

\begin{table*}
\centering
\caption{Planet engulfment parameters reproducing the observed lithium abundance. The second set of entries for each star (with stellar mass $M$ in brackets) are calculations based on a lower mass for that star (see \S\ref{engulfment})}.
\label{masses}

\begin{tabular}{l c c c c c c c c}
\hline
Target &
$A(\mathrm{Li})$ &
$M$ &
$M_{\mathrm{env}}$ &
$M_{\mathrm{inst}}$ &
$M_{\mathrm{planet}}$ &
age &
$t_{\mathrm{engulf}}$ &
$t_{1/2}$ \\
 &
observed &
$M_\odot$ &
$M_\odot$ &
$M_\oplus$ &
$M_\oplus$ &
Myr &
Myr &
Myr\\
\hline
\multicolumn{9}{l}{\textbf{NGC 2516}} \\
\addlinespace[0.5ex]
07581528-6059334 & 1.30  & 0.49   & 0.128 & 3.9 & 7.8  & 209 & 1.1      & 0.9  \\
 &      &        &         &     &  39   & 209 & 3        & 0.9 \\
            &    & (0.44) & 0.151 & 4.5 & 9.0 & 209 & $\sim$0.1 & 0.5 \\
 &      &        &         &     & 45 & 209 & 0.4      & 0.5 \\
\addlinespace[0.5ex]
07584014-6107216 & 1.87 &  0.64   & 0.084 & 9.5& 19 & 209 & 56       & 57  \\
 &      &        &         &     & 95  & 209 & 150      & 57  \\
        &       & (0.59) & 0.095 & 10.9  & 21.8   & 209 & 23       & 25  \\
 &      &        &         &     & 109  & 209 & 83       & 25  \\
\addlinespace[0.5ex]
07591796-6045189 & 1.61 & 0.59   & 0.095 & 5.9 & 11.8   & 209 & 23       & 26   \\
 &      &        &         &     & 65   & 209 & 83       & 26  \\
         &    & (0.54) & 0.111 & 6.8 & --   & --  & --       & --  \\
\addlinespace[0.5ex]
08024838-6047096 & 1.18 & 0.52   & 0.119 & 2.7 & 5.4    & 209 & 2        & 2  \\
 &      &        &         &     & 27   & 209 & 10       & 2   \\
        &    & (0.47) & 0.136 & 3.1 & --   & --  & --       & --  \\
\addlinespace[0.5ex]
\multicolumn{9}{l}{\textbf{NGC 2451A}} \\
\addlinespace[0.5ex]
07460836-3753518 & 1.39 & 0.60   & 0.201 & 7.4 & 14.8 & 50  & 1.5      & 2   \\
 &      &        &         &     & 74   & 50  & 5        & 2   \\
            &    & (0.55) & 0.233 & 8.6 & --   & --  & --       & --  \\
\addlinespace[0.5ex]
\multicolumn{9}{l}{\textbf{Blanco 1}} \\
\addlinespace[0.5ex]
00022426-3009090 & 1.47 &  0.57   & 0.089 & 3.9 & 7.8  & 110 & 16       & 26\\
 &      &        &         &     & 39   & 110 & 33       & 26  \\
            &     & (0.52) & 0.116 & 5.2 & 10.4 & 110 & 3        & 4   \\
  &      &        &         &     & 52   & 110 & 9.5      & 4   \\
\hline
\multicolumn{9}{l}{\footnotesize\textit{Notes.}} \\
\multicolumn{9}{l}{\footnotesize
$M_{\mathrm{env}}$ is the mass of the convective envelope.} \\
\multicolumn{9}{l}{\footnotesize
$M_{\mathrm{inst}}$ is the instantaneously engulfed planet mass (with Li mass fraction $1.1\times 10^{-6}$) to give $A(\mathrm{Li})$.} \\
\multicolumn{9}{l}{\footnotesize
$M_{\mathrm{planet}}$ is the modelled mass of a previously engulfed planet with a lithium mass fraction of $1.1\times 10^{-6}$.} \\
\multicolumn{9}{l}{\footnotesize \ There are two rows here, for $M_{\mathrm{planet}} = 2M_{\mathrm{inst}}$ and $M_\mathrm{planet}=10M_\mathrm{inst}$ respectively.}\\
\multicolumn{9}{l}{\footnotesize
$t_{\mathrm{engulf}}$ is how long ago such an engulfment would have taken place to produce the observed $A$(Li).} \\
\multicolumn{9}{l}{\footnotesize
$t_{1/2}$ is the time (half-life) for the observed $A(\mathrm{Li})$ to decrease by a factor of two.} \\
\end{tabular}
\end{table*}

To give an idea of these timescales we used the same \citet{Baraffe2015a} models and stellar masses used to estimate $M_{\rm inst}$, then determined how long ago the star could have accreted (instantaneously) $2\,M_{\rm inst}$ or $10\,M_{\rm inst}$ such that it would have its current observed $A$(Li). This assumes that the planetary material is mixed immediately throughout the CZ. Then, starting from the current $A$(Li) a "half-life" for the current Li enhancement was estimated. Only convective mixing is considered here, so the results, reported in Table~\ref{masses} as $t_{\rm engulf}$ and $t_{1/2}$ respectively, are formally upper limits to these timescales. Examples of these tracks, starting from the engulfment event, are shown in the top panel Fig.~\ref{lievolplot}, for two of the NGC~2516 outliers with inferred masses of $0.52M_\odot$ and $0.64M_\odot$ respectively.

A summary of these investigations is that the lowest mass outliers ($\sim 0.5M_\odot$, in NGC~2516) have the shortest $t_{\rm engulf}$ (1--10\,Myr, depending on whether $2\,M_{\rm inst}$ or $10\,M_{\rm inst}$ was accreted) and the shortest $t_{1/2}$ (1-2\,Myr) -- note the steep gradient of the example track for the $0.52M_\odot$ star in Fig.~\ref{lievolplot}. That is because their CZ bases are still hot enough to burn Li rapidly. In contrast, the highest mass outliers ($\sim 0.6M_\odot$) have CZs that are close to the Li-burning threshold temperature in these models and the timescales are about an order of magnitude longer -- the depletion track for the $0.64M_{\odot}$ star is much shallower. The important point here is that, although these higher mass stars had core temperatures high enough to burn all their Li within 20\,Myr for all the models shown in Fig.\ref{lievolplot}, the development of the radiative core reduces the CZ base temperature at later times and allows newly accreted Li to survive for longer.

Similar models were also run for the highest and lowest mass outliers in NGC~2516 and the outlier in Blanco~1 at the slightly lower masses that would be inferred from unspotted models (see Table~\ref{masses}). The small ($-0.05M_\odot$) change in mass leads to much shorter timescales. This indicates that these timescales are very sensitive to the adopted stellar parameters because the conditions at the CZ base are mass-dependent. This also suggests the timescales we have calculated will be highly model-dependent. Although we cannot do similar detailed calculations because we do not have access to the necessary models, the middle and bottom panels of Figure~\ref{lievolplot} shows how the rate of Li depletion is affected by the presence of starspots. The depletion rate is slower for spotted stars of a given mass, and for 30 per cent spot coverage Li depletion is almost halted by the development of the radiative core for $M\geq 0.6M_\odot$ at ages $>30$\,Myr. This means that Li that is engulfed after that time would likely survive to the ZAMS and beyond in those stars and that even for lower mass spotted stars the $t_{\rm engulf}$ and $t_{1/2}$ values should be larger than those in Table~\ref{masses}.

The extreme model and mass dependence of $t_{\rm engulf}$ means it is difficult to comment on how frequently engulfment events would need to occur in order to explain the small fraction of ZAMS M-dwarfs with big Li enrichments found here. If models with spots and magnetic activity are adopted or with only slightly higher inferred masses, then $t_{\rm engulf} \sim 100$\,Myr and these engulfment events would have to be quite rare -- close to the observed occurrence rate of $\sim 2-3$ per cent. On the other hand, with models that give hotter CZ bases or lower inferred masses, $t_{\rm engulf} \sim 1$\,Myr and engulfment would need to be very common or happen multiple times for a star in order to explain the small number observed; the observed occurrence rate would be a lower limit to the true frequency in this case.

\subsection{Is planetary engulfment plausible?}

The instantaneous accretion masses of 3-10$M_\oplus$ of rocky material in Table~\ref{masses} needed to explain the Li-enrichment signatures in these low-mass stars are similar to, or even less than, the masses of engulfed material invoked as an explanation for the abundance differences seen between the solar-type components of some wide binary systems  \citep{Oh2018a, YanaGalarza2021a}. The frequency of occurrence of $\geq 2$ per cent is also consistent with the inferred frequency of ingestion in those systems \citep[e.g., 3, 8 or up to 35 per cent --][respectively.]{Behmard2023b, Liu2024a, Spina2021a}

Close-in Neptune-like and larger gas giants appear to be rare around early and mid M-dwarfs -- perhaps only 0.1-0.3 per cent with periods $<10$\,d or 2-6 per cent with periods $<2$ years, about a third of the rate for solar-type hosts \citep{Quirrenbach2022a, Bryant2023a, Gan2023a, Glusman2026a}. On the other hand for lower mass close-in planets, the occurrence rates are much higher, and higher even than those of solar-type hosts in the same orbital period range -- of order 4-8  Earth or super-Earths per M-dwarf with periods $<256$\,d; 1 with periods $<50$\,d and an occurrence rate of $\sim 60$ per cent for Earth-like planets with periods $<7$\,d \citep{Dressing2015a, Hsu202a, Ment2023a}. There is therefore no shortage of rocky planetary material close to M-dwarfs, even among older field M-dwarfs.

\cite{Soares2025a} presented an analysis of detailed simulations of the formation and evolution of 1000 planetary systems around a solar-type star from the Next Generation Planetary Population Synthesis \citep[NGPPS,][]{Emsenhuber2021a}. In the NGPPS models, planetary engulfment takes place for about half the simulated systems in three broad phases: (i) During the PMS, caused by migration of giant planets through the disc. This phase ends after $\sim 10$\,Myr when the disc dissipates and we would not expect this to result in Li-enhancements because the stars have not begun Li-depletion. (ii) On the PMS and early ZAMS (the first $\sim 10-100$\,Myr), where dynamical interactions prompt collisions and scattering; some planets and planetesimals are pushed towards the star and engulfed. (iii) On longer timescales, tidal dissipation can lead to orbital decay and engulfment. The total mass accreted over the three phases has a mode of around 30-40$M_\oplus$, and the mode of each engulfment event is 3-5$M_\oplus$, easily enough to explain our Li-rich outliers. Accretion of gas giants is rare.

Of interest here are engulfments taking place in the middle phase during the PMS and ZAMS. In the NGPPS simulations, the engulfment rate (expressed per logarithmic time interval) actually peaks, and about 10 per cent of stars have at least one engulfment event, between  30-100\,Myr - the critical time for explaining any Li-enrichment. However, the simulations do suggest that dynamical interactions usually lead to smaller engulfment events ($\leq 0.1M_\oplus$). If the Li depletion rates are fast and $t_{\rm engulf}\sim 1$\,Myr, it seems unlikely that even a steady stream of small engulfments could build up an observable Li enhancement in M-dwarfs; larger engulfments than predicted by these models would be needed. On the other hand, if $t_{\rm engulf}$ is much larger then either a steady rate of small engulfments or larger, episodic events would be viable.

Initial disc mass, and to a lesser extent disc lifetime, also appear to be important in the NGPPS simulations in determining when and how much planetary material is accreted. As discussed in \S\ref{initial}, the Li-rich outliers are slowly rotating stars, which might indicate they had massive or longer-lived discs, that then favours a greater mass of accreted solids at later times. The simulations also show that the impact of accretion on observed [Fe/H] is very small; $\leq 0.07$ dex even for a $10M_\oplus$ accretion event in a sun-like star with a thin CZ. The signature would be much smaller after dilution in the larger CZ mass of M-dwarfs and unfortunately, such small metallicity enhancements would not be detectable in GES spectra, especially the complex spectra of young, active low-mass stars.

It is notable that all the outliers occur between 3550-4050 K. In the planetary engulfment scenario, the upper limit can be explained as an observational contrast effect. Stars at greater $T_{\rm eff}$ would usually still possess significant Li at the ZAMS. For example, the addition of $\sim 10M_\oplus$ of rocky material to a star already with $A$(Li)$\,\simeq 2.0$ at $T_{\rm eff} \simeq 5000$\,K (see Fig.~\ref{NGC2516Aliplot}) would increase its Li abundance by only $\sim  0.3$ dex and its EWLi from about 150\,m\AA\ to 190\,m\AA\ \citep[see fig.~A4 in][]{Jackson2025a}. Given the 1-sigma intrinsic dispersion of $\simeq 80$\,m\AA\ exhibited by K-type stars at these ages, there is no hope of unambiguously identifying Li-rich outliers with such engulfment events at these spectral types. The lack of Li-rich outliers at lower $T_{\rm eff}$, despite there being $\sim 100$ cooler stars in our sample (see Fig.~\ref{liplots}) is of marginal significance, but would be expected in the engulfment scenario. These stars have CZs deep enough to immediately mix and burn any accreted Li -- their $t_{\rm engulf}$ would be extremely short.

Another interesting observation is that there are no outliers in the three clusters older than NGC 2516. This is also of marginal significance given the sample size -- $N_M = 59$ in the clusters older than NGC~2516. If we argue that the observed fraction of Li-rich late-K and M-dwarfs is $6/259$ in the younger clusters, then an observed fraction of $0/59$ in the older clusters has a p-value of only 0.24 for the null hypothesis in a two-tailed chi-squared test. However, it could be consistent with a decreasing rate of engulfment events due to dynamical interactions post-ZAMS, combined with Li destruction timescales of $\sim 100$ Myr. It would be interesting to look for Li-rich outliers among the early M-dwarfs of well-populated older clusters like Praesepe and the Hyades.

Finally, it is fair to comment on two potential problems with the planetary engulfment scenario. The first is the peculiar distribution of EWLi among the M-dwarfs of these ZAMS clusters. The outliers were identified because they form a non-Gaussian tail to the distribution, but it is not clear why we only see isolated outliers with $1.3 < A({\rm Li}) < 2$ and $200 \leq {\rm EWLi} \leq 300$ together with completely depleted objects, rather than a more continuous distribution. Perhaps planetary engulfment events tend to be of higher mass and the Li destruction timescale short, such that objects with intermediate EWLi are rare. The curve of growth for the Li~6708\AA\ resonance line does saturate and flatten somewhat at $A$(Li)$\geq 2$ (EWLi $\geq 200$\,m\AA) in M-dwarfs, meaning there could be a modest ``pile-up" of objects at that value as Li is depleted. At present the small number of identified outliers prevents a more definitive analysis of how remarkable the distribution might be. Secondly, the NGPPS simulations analysed by \cite{Soares2025a} suggest that although the total mass engulfed during the stellar lifetime is frequently $>10M_\oplus$, not enough mass is accreted during the dynamical phase at 10--100\,Myr. However, these simulations were for solar-type stars and there is ample evidence that these might not be representative of the discs and exoplanetary architecture around lower mass stars, so more specific simulations might be required.

\section{Summary}
\label{summary}

We have searched for examples of anomalously Li-rich cool stars amongst members of open clusters that were observed as part of the {\it Gaia}-ESO spectroscopic survey. Six examples of ZAMS M-dwarfs with $3560 \leq T_{\rm eff}/K \leq 4045$ and $1.3 \leq A({\rm Li}) < 2$ \ have been found in three clusters (NGC~2516, Blanco~1 and NGC~2451a), amounting to about 2-3 per cent of the population at those spectral types. They stand out as clear outliers among their Li-depleted siblings.

The Li-rich outliers are indistinguishable from other cluster members in terms of their position, kinematics (they have membership probabilities $>0.999$), parallax and location in colour-magnitude diagrams, and 5 of them show no evidence for binarity. Their positions in the absolute colour-magnitude diagram indicate they cannot be as young as suggested by their Li content.
 
Three explanations for these outliers are considered. Inhibited PMS Li depletion, perhaps due to strong magnetic activity or starspots, is thought unlikely because the outliers are among the slowest rotating stars of their respective clusters. Odd initial conditions or accretion of pristine material from a disc also seems unlikely because Li depletion should still be very effective in these low-mass stars long after any disc has dissipated and should erase any formation peculiarities.

Instead, it seems plausible that the outliers are direct evidence of planetary engulfment -- a scenario predicted in models by \cite{Sevilla2022a}. If the engulfments take place after the development of a substantial radiative core, then the convection zone base may be sufficiently cool to allow Li survival in the envelope and at the photosphere. Instantaneous engulfed masses of $3-10\,M_\oplus$ of rocky, volatile-depleted material are needed to explain the observed $A$(Li), with only modest uncertainties and model-dependence. However, the survival time of the ingested Li is highly dependent on the inferred mass of the objects and which models are used to estimate those Li survival timescales. This means that the observed frequency of the phenomenon (2-3 per cent) may represent the true frequency of engulfment if Li survival times are long, but could be a lower limit if they are shorter.

Close-in Earth-like and Super-Earth exoplanets are common around M-dwarfs and the required planetary engulfment masses are similar to those proposed in the contested interpretation of small chemical abundance differences in the solar-type components of some binary systems. It may be that total Li-depletion, which is the normal outcome of PMS evolution in the lower mass stars of open clusters, provides a blank canvas upon which the frequency and timing of the engulfment of fresh planetary material can be assessed.

\section*{Acknowledgments}
This work has made use of data from the European Space Agency (ESA) mission
{\it Gaia} (\url{https://www.cosmos.esa.int/gaia}), processed by the {\it Gaia}
Data Processing and Analysis Consortium (DPAC,
\url{https://www.cosmos.esa.int/web/gaia/dpac/consortium}). Funding for the DPAC
has been provided by national institutions, in particular the institutions
participating in the {\it Gaia} Multilateral Agreement.

Giraffe spectra were obtained from the ESO Science Archive Facility with DOI(s): \url{https://doi.eso.org/10.18727/archive27}.

This research has made use of the SIMBAD database, operated at Centre de Donn\'ees astronomiques de Strasbourg (CDS), France.

This publication makes use of data products from the Two Micron All Sky Survey, which is a joint project of the University of Massachusetts and the Infrared Processing and Analysis Center/California Institute of Technology, funded by the National Aeronautics and Space Administration and the National Science Foundation.

The VISTA Hemisphere Survey data products served at Astro Data Lab are based on observations collected at the European Organisation for Astronomical Research in the Southern Hemisphere under ESO programme 179.A-2010, and/or data products created thereof.

This publication makes use of VOSA, developed under the Spanish Virtual Observatory (https://svo.cab.inta-csic.es) project funded by MCIN/AEI/10.13039/501100011033/ through grant PID2020-112949GB-I00.

RDJ is partly supported by Science Technology Facilities Council (STFC) grant ST/Y002407/1. IB is partly supported by STFC grant ST/Y002164/1.

\section*{Data Availability Statement}
The data used in this paper are all publicly available from: the cited publications in the case of spectral parameters and cluster membership; from CDS: \url{https://cdsarc.cds.unistra.fr/viz-bin/cat/I/355} in the case of Gaia DR3 data; and  from the ESO phase 3 archive for the GES spectra and stellar parameter catalogue: \url{https://archive.eso.org/wdb/wdb/adp/phase3_main/query?prog_id=073.D-0587}

For the purposes of open access, the author has applied a Creative Commons Attribution (CC-BY) licence to any Accepted Author Manuscript version arising from this submission.

\bibliographystyle{mnras} 
\bibliography{references}

\appendix

\section{Diagnostic plots}
\label{App1}

\begin{figure*}
    \centering
    \includegraphics[width=0.95\textwidth]{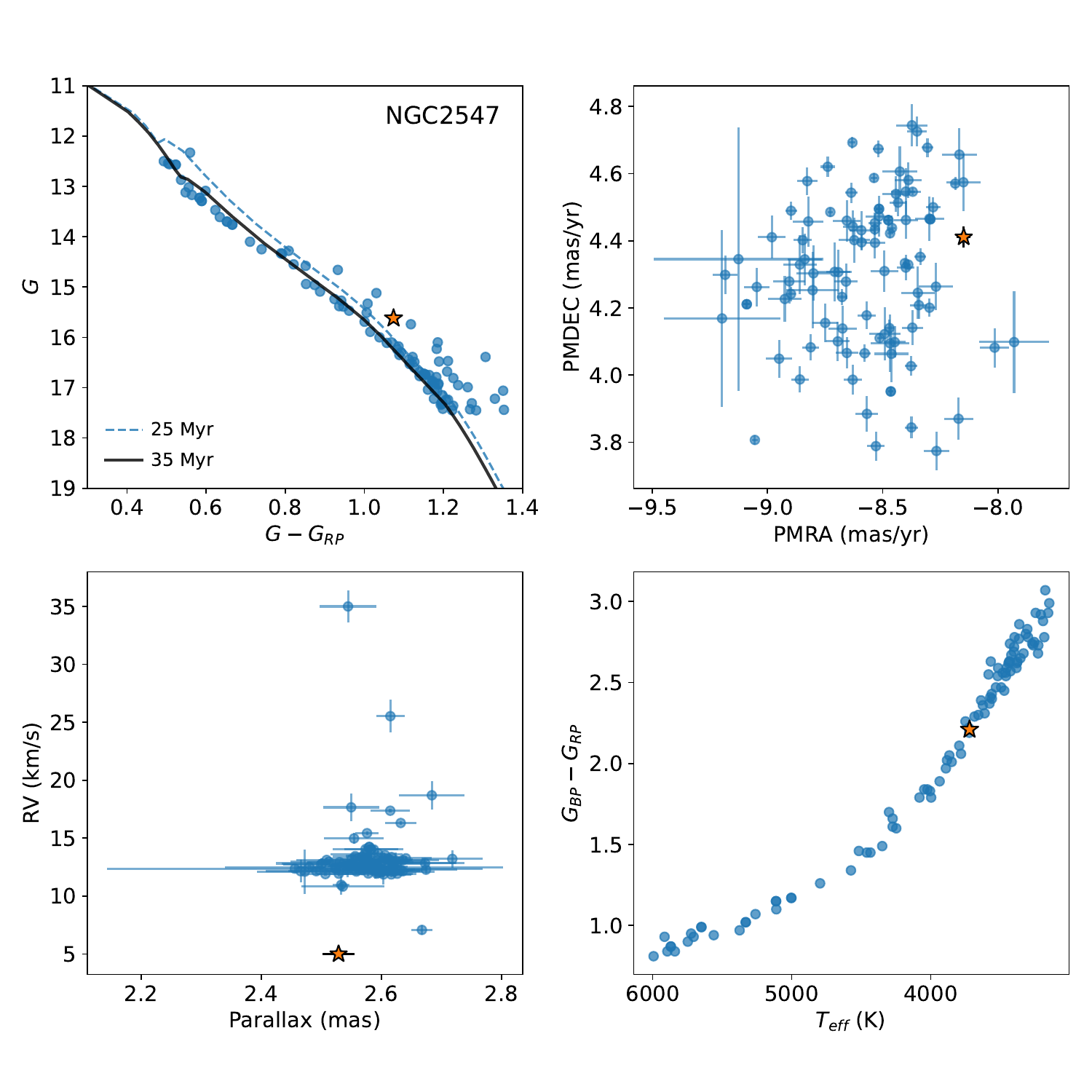}
\caption{The position of the Li-rich outliers in NGC~2547 compared with other cluster members in several diagnostic diagrams. The top-left plot shows isochrones from the {\sc spots} models of \protect\cite{Somers2020a} at the nominal age of the cluster and at 25\,Myr (i.e. where a star of age 25\,Myr but at the same distance as the cluster would appear).}
\label{NGC2547plot}
\end{figure*}

\begin{figure*}
    \centering
    \includegraphics[width=0.95\textwidth]{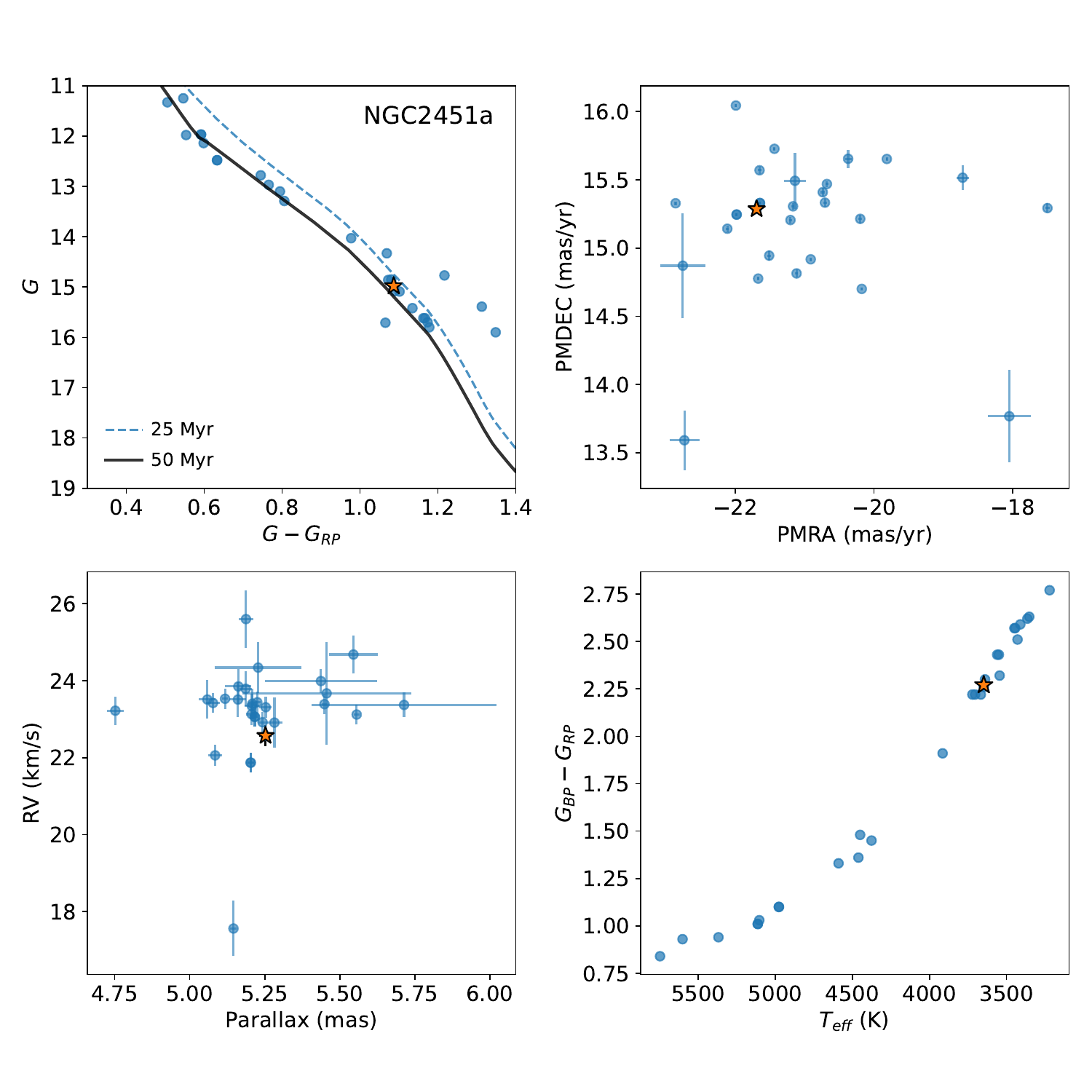}
\caption{The position of the Li-rich outlier in NGC~2451a compared with other cluster members in several diagnostic diagrams. The top-left plot shows isochrones from the {\sc spots} models of \protect\cite{Somers2020a} at the nominal age of the cluster and at 25\,Myr. (i.e. where a star of age 25\,Myr but at the same distance as the cluster would appear) }
\label{NGC2451aplot}
\end{figure*}

\begin{figure*}
    \centering
    \includegraphics[width=0.95\textwidth]{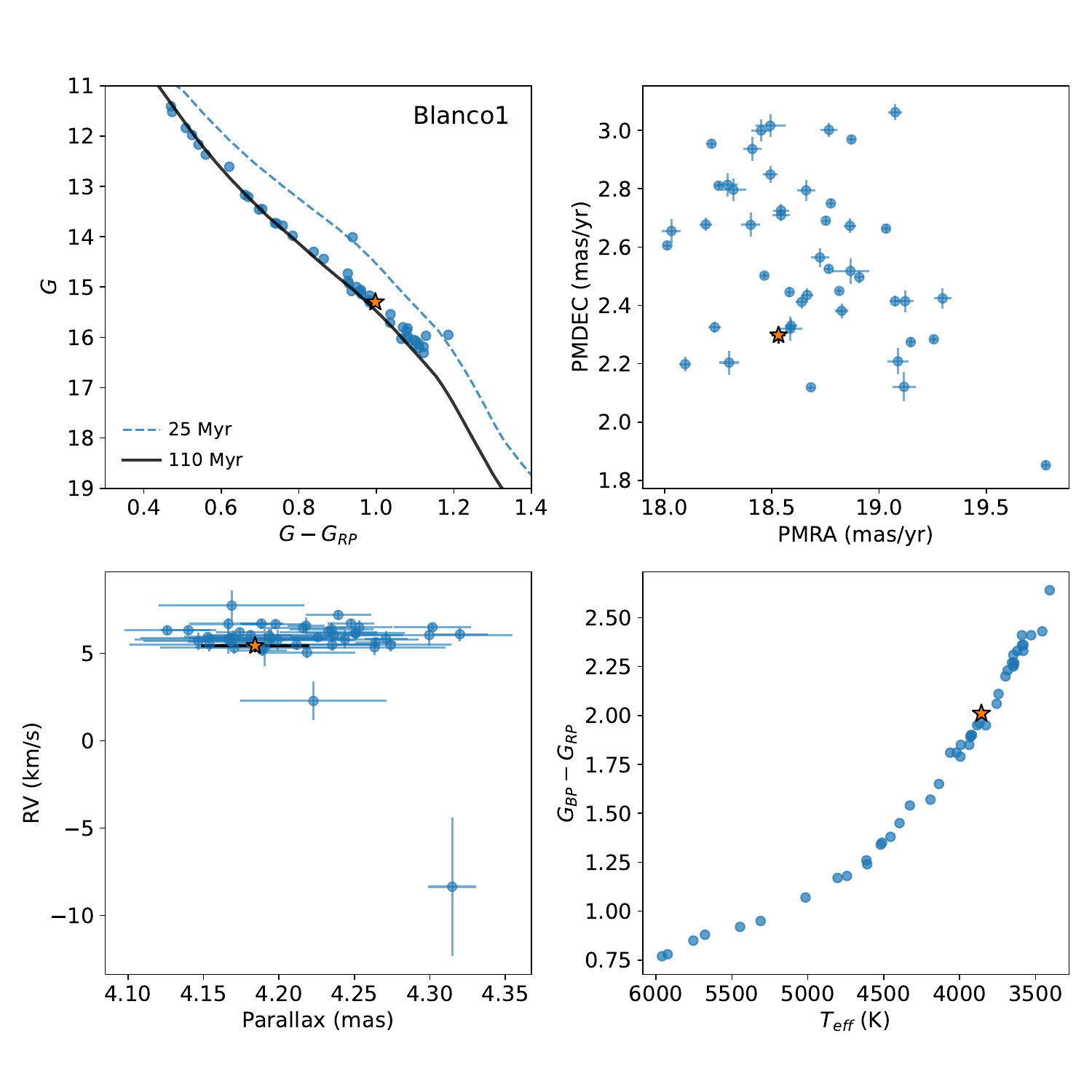}
\caption{The position of the Li-rich outlier in Blanco~1 compared with other cluster members in several diagnostic diagrams. The top-left plot shows isochrones from the {\sc spots} models of \protect\cite{Somers2020a} at the nominal age of the cluster and at 25\,Myr (i.e. where a star of age 25\,Myr but at the same distance as the cluster would appear).}
\label{Blanco1plot}
\end{figure*}

Diagnostic plots for the outliers in NGC~2547, NGC~2451a and Blanco~1, similar to that for NGC~2516 (see Fig.~\ref{NGC2516plot}) are shown in Figs.~\ref{NGC2547plot}, \ref{NGC2451aplot} and \ref{Blanco1plot}.

\section{Spectral Energy Distribution fitting}

\label{AppSED}

The spectral energy distributions (SEDs) of the Li-rich outliers were modelled in order to check the spectroscopic estimates of $T_{\rm eff}$ and assumed extinction values.

The observed SEDs were constructed using Gaia DR3 ($G$, $G_{BP}$, $G_{RP}$, 2MASS \citep[$JHK_s$,][]{Skrutskie2006a} and Vista Hemisphere Survey \citep[$JK_s$,][]{McMahon2013a}.
The SEDs were modelled using the {\sc bt-settl cifist} atmospheres \citep{Allard2012a} implemented as part of the ``Virtual Observatory SED Analyzer" (VOSA) tool \citep{Bayo2008a}. The free parameters were $T_{\rm eff}$ and the extinction, whilst $\log g$ was fixed at 4.5 and the metallicity at the solar composition \citep[defined by][]{Caffau2011a}. The Bayes analysis fitting mode was used.

All of the SEDs were well-modelled by the chosen atmospheres, with no sign of any near-IR excess. The $T_{\rm eff}$ values are reported (as $T_{\rm SED}$) in Table~\ref{outliers} and are precise to within the 100\,K spacing of the model grid. The returned extinction ($A_V$) estimates were within 0.05 mag of those estimated for their clusters by \cite{Jackson2022a}.

\section{Lithium abundance plots}
\label{AppALi}

\begin{figure}
    \centering
    \includegraphics[width=0.48\textwidth]{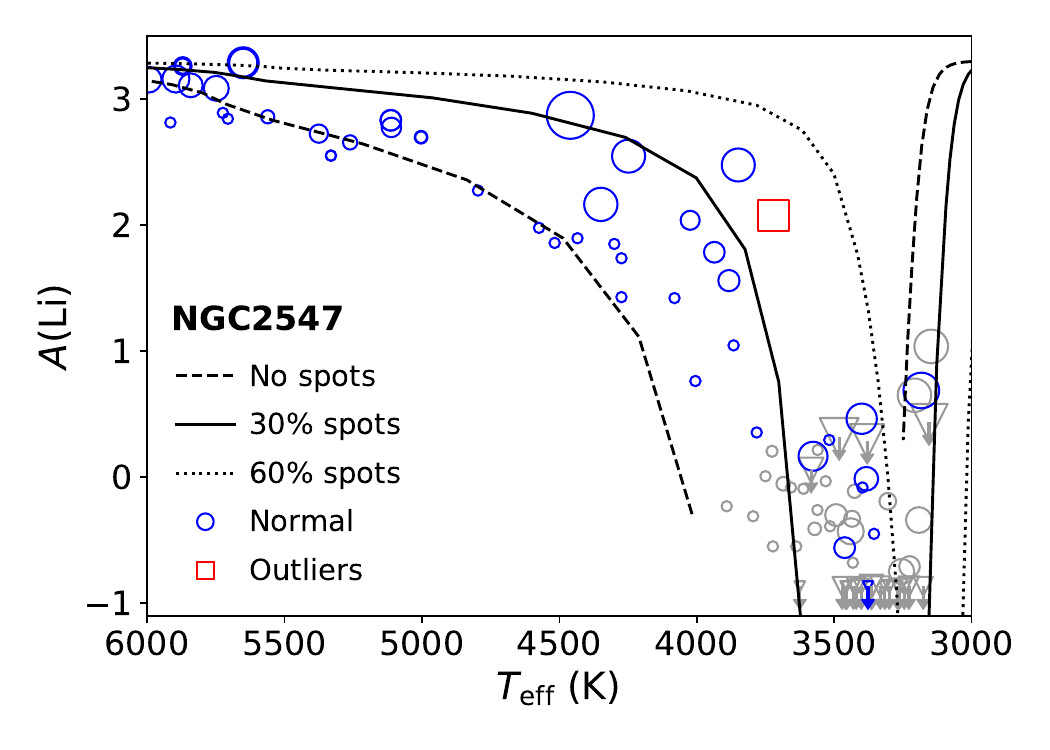}
    \includegraphics[width=0.48\textwidth]{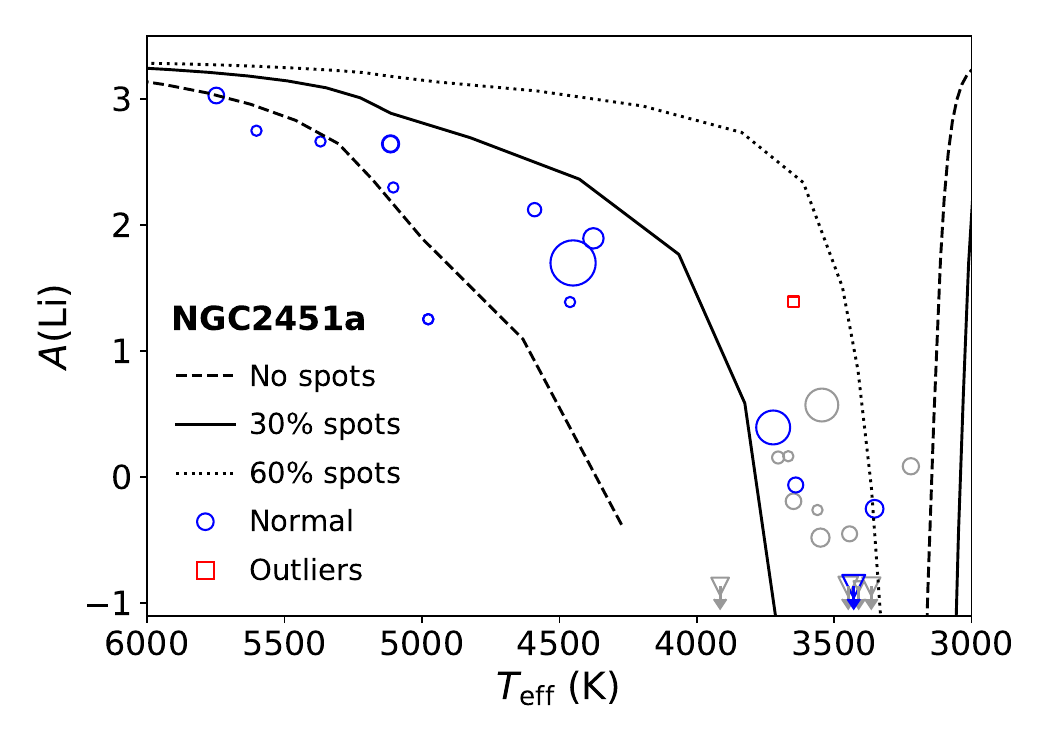}
    \includegraphics[width=0.48\textwidth]{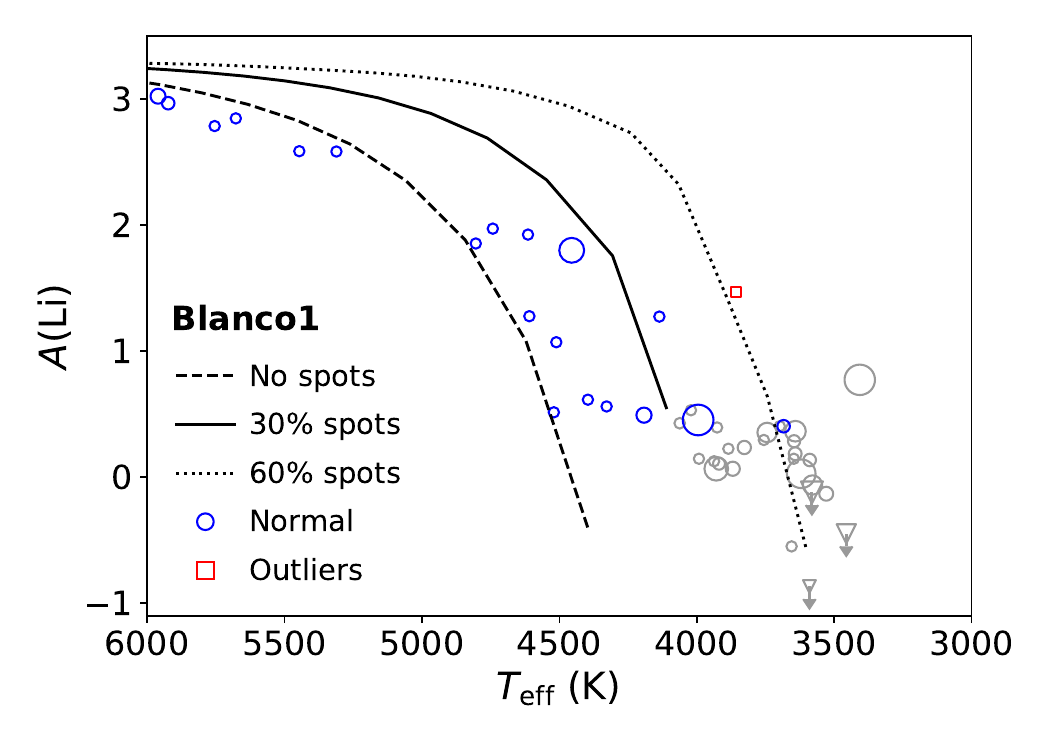}
\caption{The position of the Li-rich outliers in NGC~2547, NGC~2451a and Blanco~1 in an $A$(Li) versus $T_{\rm eff}$ diagram. The observational data are from the GES DR4 abundance catalogue. The size of the points is scaled with $v \sin i$ as in Fig.~\ref{liplots}. The points shaded in grey are those which have an EWLi smaller than twice its error bar and which arguably should also be treated as upper limits (see \S\ref{inhibit}). Isochrones of Li depletion from the {\sc spots} models of \protect\cite{Somers2020a}, at the nominal age of the cluster, are shown for spot areal coverage fractions of 0, 30 and 60 per cent.}
\label{AppALiplot}
\end{figure}

Figure~\ref{AppALiplot} shows the NLTE Li abundances versus $T_{\rm eff}$ for stars in NGC~2547, NGC2451a and Blanco~1 compared with {\sc spots} evolutionary models.

\bsp 
\label{lastpage}
\end{document}